\documentclass[preprint]{aastex}
\newcommand{\kmps}{km~s$^{-1}$} 
\newcommand{\phoe}{{\tt PHOENIX}}
\newcommand{\snia}{SN~Ia} 
\newcommand{\sneia}{SNe~Ia}

\newcommand{\snnfd}{SN~1994D}
\newcommand{\umb}{\ensuremath{U\!-\!B}}
\newcommand{\bmv}{\ensuremath{B\!-\!V}}
\newcommand{\vmr}{\ensuremath{V\!-\!R}}
\newcommand{\rmi}{\ensuremath{R\!-\!I}}

\newcommand{\richm}{\tablenotemark{b}}
\newcommand{\patm}{\tablenotemark{c}}
\newcommand{\intm}{\tablenotemark{e}}
\newcommand{\jktm}{\tablenotemark{d}}
\newcommand{\siii}{\ion{Si}{2}}
\newcommand{\feii}{\ion{Fe}{2}}
\newcommand{\cahk}{\ion{Ca}{2}~H+K}
\newcommand{\cair}{\ion{Ca}{2}~IR triplet}
\newcommand{\mgii}{\ion{Mg}{2}}
\newcommand{\hei}{\ion{He}{1}}
\newcommand{\ebmv}{E(B-V)}
\newcommand{\siiw}{\ion{S}{2}~``W"}
\newcommand{\naid}{\ion{Na}{1}~D}
\newcommand{\tstd}{\ensuremath{\tau_{5000}}}
\newcommand{\nifs}{$^{56}$Ni}
\newcommand{\gamray}{$\gamma$-ray}

\usepackage{psfig}
\begin{document}

\title{NLTE Synthetic Spectral Fits to the Type~I\lowercase{a} 
Supernova~1994D in NGC~4526}

\author{Eric~J. Lentz\altaffilmark{1}, E.~Baron, David Branch} 
\affil{Department of Physics and Astronomy, University of Oklahoma, 
440 W. Brooks St., Norman, OK 73019-0225} 
\email{lentz, baron, branch@mail.nhn.ou.edu}
\author{and} 
\author{Peter~H. Hauschildt} 
\affil{Department of Physics and Astronomy \& Center for Simulational 
Physics, University of Georgia, Athens, GA 30602-2451} 
\email{yeti@hal.physast.uga.edu}

\altaffiltext{1}{Present address: Department of Physics and Astronomy
\& Center for 
Simulational Physics, 
University of Georgia, Athens, GA 30602}

\begin{abstract}

We have fit the normal, well observed, Type Ia Supernova (\snia)
\snnfd\ with non-LTE spectra of the deflagration model W7.  We find
that well before maximum
luminosity W7 fits the optical spectra of \snnfd.  After maximum
brightness the quality of the fits weakens 
as the spectrum forms in a core rich in iron peak elements.  We show the
basic structure of W7 is likely to be representative of the typical \snia. We
have shown that like W7, the typical \snia\ has a layer of unburned
C+O composition at $v > 15000$ \kmps, followed by layers of C-burned
and O-burned material with a density structure similar to W7.  We
present UVOIR (UBVRIJKH) synthetic photometry and colors and compare with
observation. We have computed the distance to the host galaxy,
NGC~4526, obtaining a distance modulus of $\mu = 30.8 \pm 0.3$.  We
discuss further application of this direct measurement of \sneia\
distances.  We also discuss some simple modifications to W7 that could
improve the quality of the fits to the observations.

\end{abstract} 
\keywords{galaxies: individual (NGC~4526) --- stars: atmospheres ---
supernovae: individual (SN~1994D)} 
\clearpage
\section{Introduction}

Type Ia Supernovae (\sneia) have contributed to the chemical evolution
of the universe by releasing iron peak and alpha chain elements into
the various star forming and other gaseous regions.  As \sneia\ are
among the brightest objects known and their light curves can be used
to improve their uniformity, they are among the most useful objects
with which to study the structure
and nature of the universe \citep{riess_scoop98,perletal99}. It is
therefore important to have a fuller 
understanding of the underlying physics and origin of \sneia.

The current understanding of the progenitor system for \sneia\
features a sub-Chandrasekhar mass C+O white dwarf accreting mass from
an ordinary star until reaching the Chandrasekhar mass, or by merging
with another C+O white dwarf and then exploding \citep[cf.][and
references therein]{prog95,livio99}.  \sneia\ scenarios
with sub-Chandrasekhar mass  explosions have not
been able to reproduce the observed spectra and photometry
\citep{nughydro97,hofasi97}.

SN 1994D, in NGC 4526, was discovered 2 weeks before maximum
brightness \citep{iauc94d}.  It was one of the best observed \sneia,
with near-daily spectra starting 12 days before maximum brightness
(-12 days) and continuing throughout the photospheric phase.  \snnfd\
has been well observed photometrically
\citep{richsn94d,patat94D96,meikle94d91t,tsv94d94i} and
spectroscopically
\citep{filipasi97,patat94D96,meikle94d91t}. \citet{wang97} found no
significant polarization in \snnfd\ 10 days before maximum light.
\citet{cum94d96} placed a limit on a solar-composition progenitor wind
of $1.5 \times 10^{-5}$~M$_{\odot}$~yr$^{-1}$ for a 10 \kmps\ wind.
\snnfd\ has been previously modeled with synthetic spectra and light
curves by several groups
\citep{hatano94D99,hofsn94d,meikle94d91t,mazz94d}. Our goal in this
paper is to take a large set of observed spectra and a detailed (even
if somewhat parameterized) hydrodynamical model and confront the
two. \citet{hofsn94d} did this to some extent, but focused on the
photometry more heavily, partly because much of the spectroscopy has 

\sneia\ with spectral coverage this frequent and early are still quite
rare.  Other \sneia\  with excellent spectral coverage are: SN~1989B
\citep{wellsetal89B}, discovered approximately 7 days before maximum
light,  SN~1996X
\citep{salvoetal01}, discovered 6 days before maximum light, and
SN~1998bu \citep{jhaetal98bu99,hern98bu00}, discovered 10 days before
maximum light.   Both of these well observed, normal \sneia\
were discovered after the phase in \snnfd\ when daily spectral
coverage had already begun.  This aspect of the available data for
\snnfd\ make it still the best candidate for detailed, multi-epoch
spectrum synthesis of the photospheric phase, especially at the
earliest dates.  This detailed and early set of spectra allow us to
probe the spectrum formation at early epochs, and therefore the
outermost layers of the supernova.

\section{Numerical Technique}

We have used the multi-purpose spectrum synthesis and model atmosphere
code \phoe~{\tt 9.1}\ \citep[see][and references
therein]{hbjcam99}. \phoe\ has been designed to accurately include the
various effects of special relativity important in rapidly expanding
atmospheres, like supernovae.  Ionization by non-thermal electrons
from \gamray s from the nuclear decay of \nifs\ that powers the light
curves of \sneia\ is taken into account.  We have tested, in a few
models, an updated method for calculating the \gamray\ deposition
using a solution of the spherically symmetric radiative transfer
equation for \gamray s with \phoe, and found no difference in
deposition or temperature convergence relative to the older, single
$\Lambda$-iteration technique of \citet{nugphd} which is based on the
method of \citet{sw84}.  We used an effective \gamray\ opacity,
$\kappa_{\gamma} = 0.06$~Y$_e$~cm${^2}$~gm$^{-1}$ \citep{cpk80} for all
calculations.  In the models presented here we solve the NLTE rate
equations for \ion{H}{1}~(10 levels/37 transitions),
\ion{He}{1}~(19/37), \ion{He}{2}(10/45), \ion{C}{1}~(228/1387),
\ion{O}{1}~(36/66), \ion{Ne}{1}~(26/37), \ion{Na}{1}~(3/2),
\ion{Mg}{2}~(73/340), \ion{Si}{2}~(93/436), \ion{S}{2}~(84/444),
\ion{Ti}{2}~(204/2399), \ion{Fe}{2}~(617/13675), and
\ion{Co}{2}~(255/2725), which we have chosen both for computational
efficiency and to be consistent with what was used in our earlier work
\citep{nughydro97,nug1a95}. In future work we will use more species in
NLTE, but test calculations show that this is the primary feature
forming species of each element and that it  set is quite adequate.

The model W7 was prepared by homologous expansion using a rise time of
20 days after explosion to maximum light
\citep[e.g.,][]{riessetalIaev00,akn00}. The hydrodynamic output was
extended from $\sim 24000$~\kmps\ to 30000~\kmps\ with the unburned C+O
white dwarf composition as in previous \phoe\ calculations using W7
\citep{nughydro97,nugphd,lentzmet00}.
At each epoch, we have fit the luminosity to match the shape and color
of the observations, while solving for the energy balance and converged
NLTE rate equations. To fit the spectra, after having fixed the date of
explosion (20 days before maximum brightness in B), the model (W7), and
the NLTE species, the only parameter we allow to vary is the bolometric
luminosity of the model. The bolometric luminosity for a model with a
fixed density structure is a required boundary condition for the
system of equations, but it may also be convienently thought of in
terms of a ``model temperature'' so that one can see that varying the
model temperature is the degree of freedom necessary to 
match of the shape of the synthetic spectrum to the observations.
We have found that with luminosity changes of less than
20\% it is difficult to discern differences in the quality of the fit,
except in certain cases with good observed spectral
coverage and good synthetic spectral fits. This corresponds to about
5\% differences in 
temperature and about 0.2 magnitudes in the absolute luminosity
calibration. 
Our selection of the best fit is done by eye and at least two of the
authors independently select the best fit. While a lot of experience
is involved in selecting the best fit, we first strive to fit the
overall shape (or colors) and then we strive to fit the lineshape of
selected lines.
We are developing statistical tests to improve the
sensitivity and we note that we are sensitive to much smaller
variations of order 
0.02 mag when this method is applied to SNe~II (R.~Mitchell et al., E.~Baron
et al., in preparation).

\section{Synthetic Spectra\label{spec}}

We have calculated synthetic spectra from -12 days after maximum light
(day 8 after explosion) to 12 days after maximum light (day 32 after
explosion). The plotted synthetic spectra have been multiplied by an
arbitrary constant to give the best fit to the de-redshifted and
de-reddened observation.  We have adopted the reddening, $\ebmv =
0.06$, of \citet{patat94D96}. The quoted error, $\pm 0.02$, is
consistent with the other estimates of $\ebmv$ \citep[cf.][]{DR94D99}.
Using $\ebmv = 0.06$, we have applied the \citet{card89} reddening law
to deredden the observations for comparison with the synthetic
spectra. We have de-redshifted the observed spectra by
$v=830$~\kmps\ \citep{cum94d96} for the heliocentric velocity of the
supernova, which is at the 0.3~\% level.  Synthetic optical and
infra-red photometry are listed in Table~\ref{tab:phot}.  Observed
colors and reddened synthetic colors are listed in
Table~\ref{tab:color}\ for comparison.

\subsection{Before Visual Maximum}

\subsubsection{9 March 1994\label{mar9}}
\begin{figure}
\begin{center}
\psfig{file=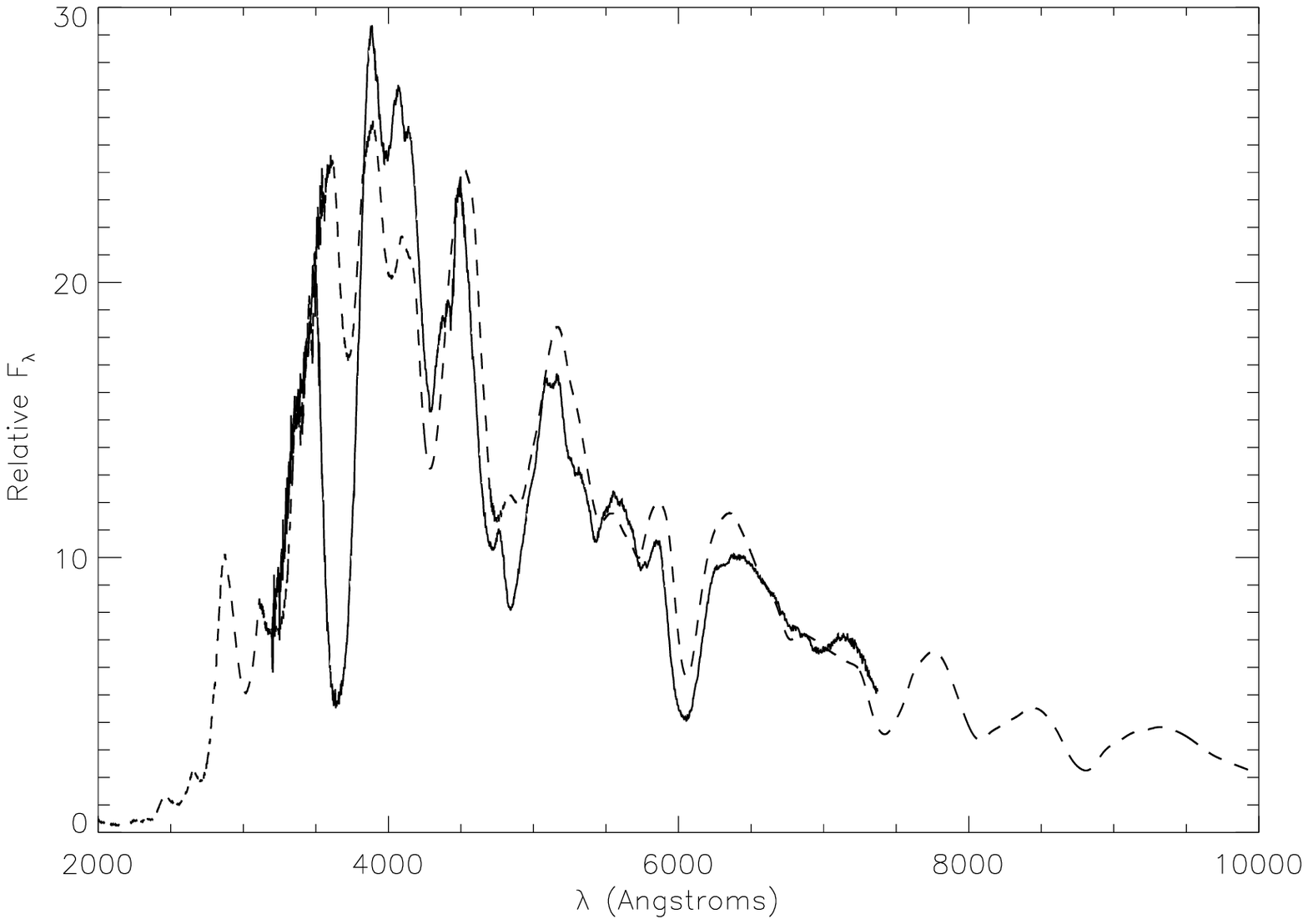,width=14cm}
\caption{\snnfd\ on 9~March 1994 
\protect\citep[solid line,][]{filipasi97} and W7 best fit synthetic spectrum for 
day 8 after explosion (dashed line).\label{fig:d08}}
\end{center}
\end{figure}

We plot the 9~March 1994 observed spectrum of \snnfd\ with our best
synthetic spectrum at day~8 after the explosion in
Figure~\ref{fig:d08}.  The features of the synthetic spectrum match
the observations in strength and shape, except the large \cahk\
absorption at 3800~\AA.  This feature is not well produced in the
earliest synthetic spectra due to over-ionization, which we discuss in
\S~\ref{cana}.  The 6000~\AA\ \siii\ absorption is shifted upward
somewhat by a small difference in the `continuum' between the model
and \snnfd, but the strength and shape are otherwise fine. Of special
note is the broad \feii\ feature at
4500--5000~\AA. \citet{hatano94D99} attribute this feature to
high-velocity iron in the unburned, or at least unenhanced by fresh
iron, layers of the supernova. We concur with this conclusion, and
note that if any minima exist in the \feii\ opacity, as they needed to
fit the shape the feature, the minima arise naturally from the
radiative equilibrium and NLTE treatment within the W7 model.  The red
minimum in the absorption is weaker in our fit than in the
observation, but the two minima are clearly indicated. This feature is
robust. In a model with one-half the luminosity, the \feii\ feature
remains strong and continues to form in high-velocity material, though
the total flux blueward of that feature (synthetic spectrum not
plotted) is greatly reduced relative to the red flux. We have also
tested the \ion{Fe}{2}\ hypothesis. In a spectrum containing only
\ion{Fe}{2}\ line opacity the feature appeared with the same shape and
strength.  This technique is discussed in \S~\ref{ir105}.

The overall shape of the synthetic spectrum fits the peaks in the
spectrum as well as the `continuum' flux level blue of the
\ion{Ca}{2}~H+K feature and red of the \ion{Si}{2}\ feature. The \bmv\
and \vmr\ colors are within 0.1~mag of the observed values, which is
less than the estimated 
errors, and therefore in agreement.  The weakness of the \cahk\ feature in
the synthetic spectrum decreases (brightens) $M_U$ by about 0.2~magnitudes.

\subsubsection{10 March 1994}

\begin{figure}
\begin{center}
\psfig{file=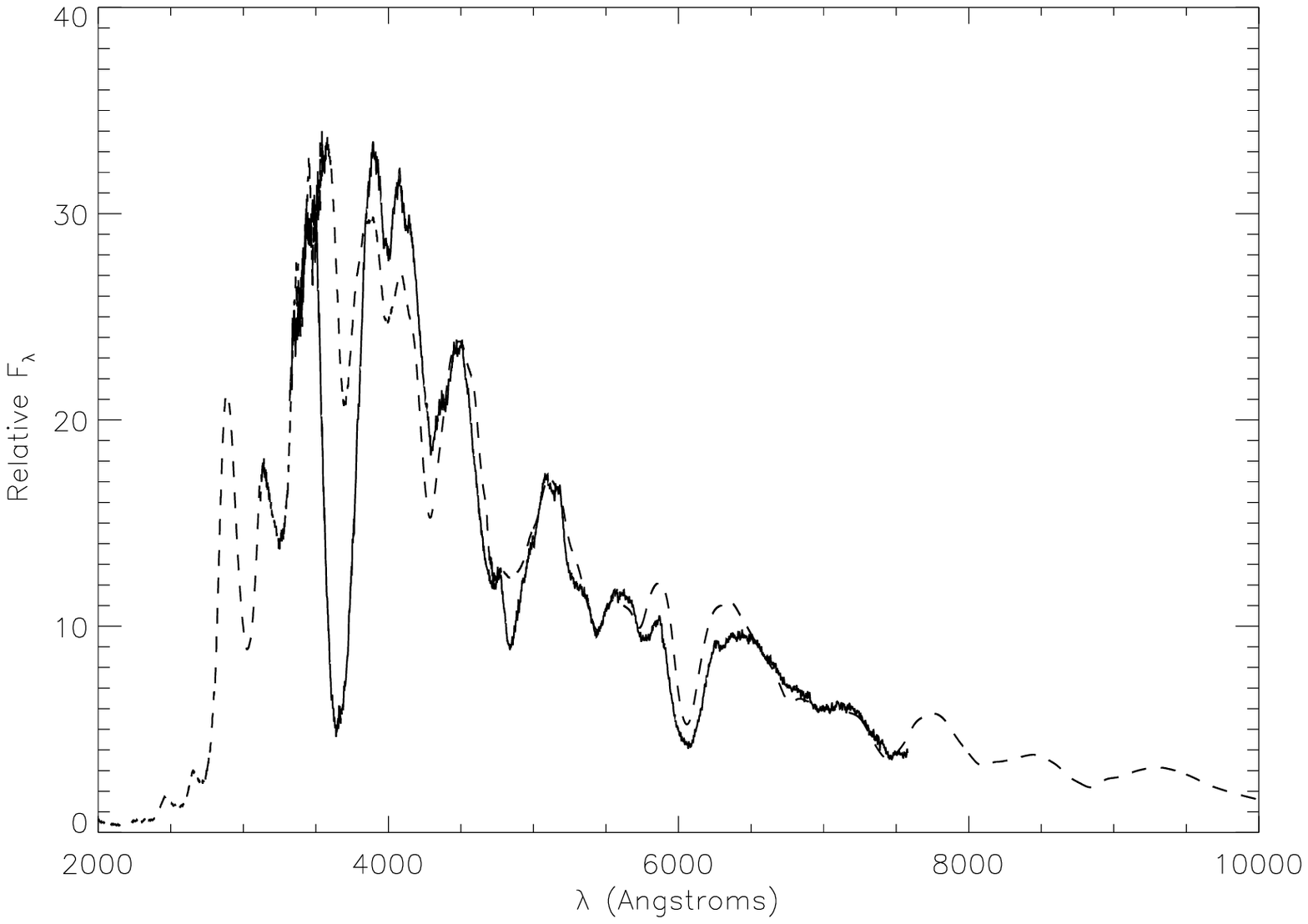,width=14cm}
\caption{\snnfd\ on 10~March 1994 \protect\citep[solid
line,][]{filipasi97} and W7 best fit synthetic spectrum for day 9
after explosion (dashed line).\label{fig:d09}}
\end{center}
\end{figure}

We plot the 10~March 1994 observed spectrum of \snnfd\ with our best
synthetic spectrum at day~9 in Figure~\ref{fig:d09}. The `continuum'
levels and spectral features of the synthetic again match the
observation except for the \cahk\ absorption. The features are
similar to the 9~March spectrum and the day~8 synthetic spectrum,
including the deep, fast \ion{Fe}{2}\ feature. The colors are
consistent with observations except \umb, which is too small, clearly
caused by the weak \cahk\ feature in the synthetic spectrum.

\subsubsection{11 March 1994}

\begin{figure}
\begin{center}
\psfig{file=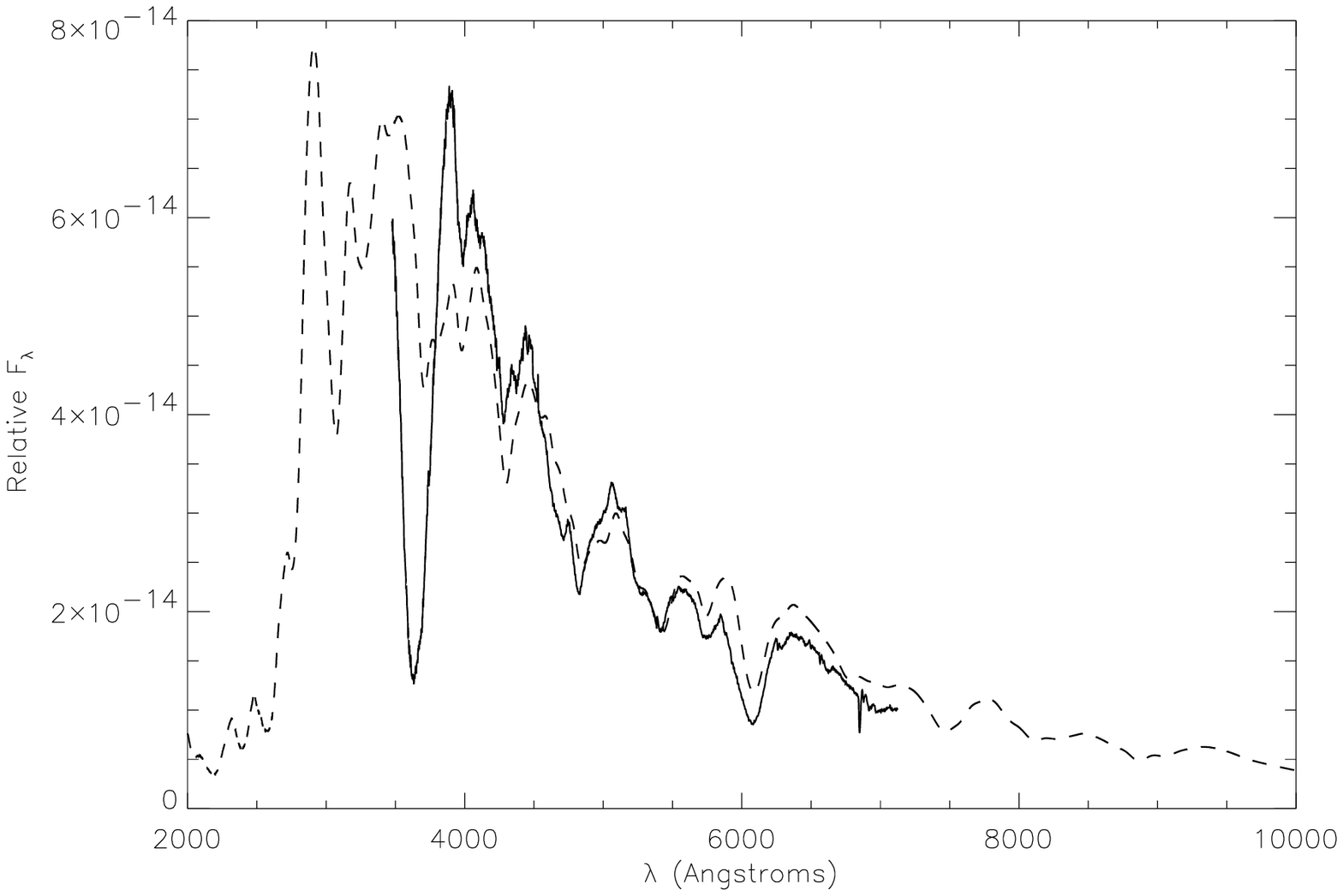,width=14cm}
\caption{\snnfd\ on 11~March 1994 \protect\citep[solid line,][]{patat94D96}
and W7 best fit synthetic spectrum for
day 10 after explosion (dashed line).\label{fig:d10}}
\end{center}
\end{figure}

We plot the 11~March 1994 observed spectrum of \snnfd\ with our best
synthetic spectrum at day~10 in Figure~\ref{fig:d10}. Note that the
observations obtained by Patat are in cgs, whereas those obtained at
Lick are in arbitrary units. Both scales are linear, so the reader
should be able to judge the fit quality equally well when comparing
between spectra obtained by different observers. Again the model
and observation are quite similar to the previous two dates. However,
the \feii\ feature is changing in shape and is no longer as
strong. This is reflected in the synthetic spectrum. The synthetic
spectrum again falls short of the peak of the observed flux near
4000~\AA, but this would be compensated if the \cahk\ feature were
stronger. This model has been chosen for the fit to the strength and
shape of the \siii\ features and our estimate of the flux from the
previous observations blueward of the \cahk\ feature. The synthetic
and observed optical colors are consistent. A good fraction of the
0.4~mag difference in the \umb\ colors is related to the weak \cahk\
feature.

\subsubsection{12 March 1994}

\begin{figure}
\begin{center}
\psfig{file=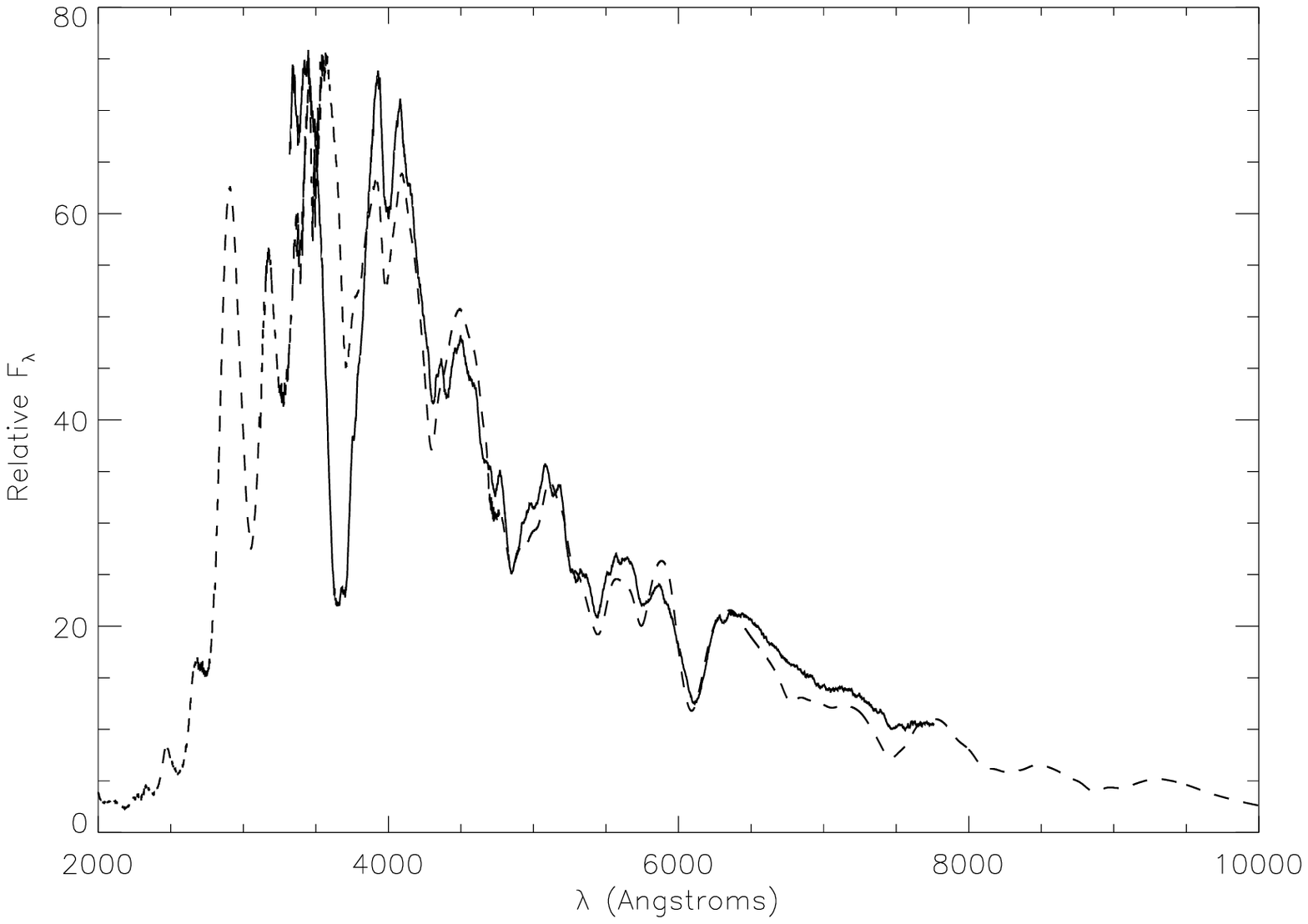,width=14cm}
\caption{\snnfd\ on 12~March 1994 \protect\citep[solid line,][]{filipasi97}
and W7 best fit synthetic spectrum for
day 11 after explosion (dashed line).\label{fig:d11}}
\end{center}

\end{figure}

The 12~March 1994 observed spectrum of \snnfd\ and our best synthetic
spectrum at day~11 are displayed in Figure~\ref{fig:d11}.  Most of the
features in the synthetic spectrum fit quite well at this epoch.  The
\cahk\ feature is still weaker than the observed, but the relative
strength has improved over the previous day. This is reflected in the
better match of the \umb\ synthetic color to the two observations.
The fact that the model is bluer than the observations to the red of
6000~\AA\ is likely due to the fact that there is not enough line
blanketing produced by W7, (which would push flux from the blue to the
red). We shall see that this is likely due to a flaw in the density
and/or composition structure of the inner parts of W7.  The remaining
colors are consistent with observations.

\subsubsection{13 March 1994}

\begin{figure}
\begin{center}
\psfig{file=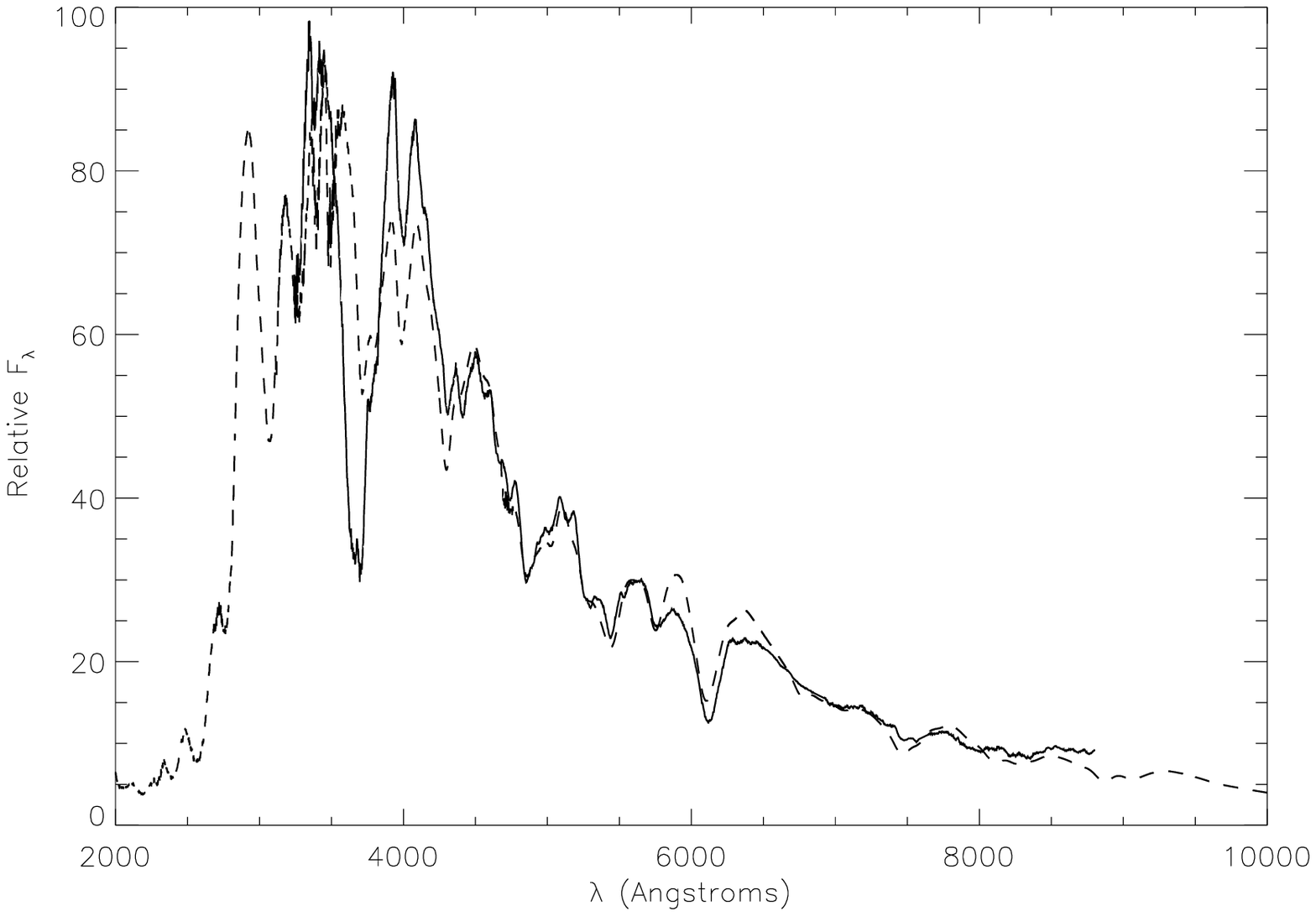,width=14cm}
\caption{\snnfd\ on 13~March 1994 \protect\citep[solid line,][]{filipasi97}
and W7 best fit synthetic spectrum for
day 12 after explosion (dashed line).\label{fig:d12}}
\end{center}
\end{figure}

Figure~\ref{fig:d12} shows the 13~March 1994 observed spectrum of
\snnfd\ with our best 
synthetic spectrum at day~12.  The synthetic
spectrum fits the observation very well, with nearly all features
represented, although some strengths are incorrect.  The \cahk\
feature is still weaker than the observation, but it now shows the
`split' which we have previously shown to be generated by a blend of
\cahk\ and \siii~$\lambda~3858$ \citep{nughydro97,lentzmet00}. The
blue \siii\ feature at 5800~\AA\ is stronger than in the
observation. \citet{nughydro97}\ have shown that the relative strength
of the two features are an effective temperature diagnostic. The
stronger blue \siii\ feature in the model indicates that the \siii\
line-forming region is cooler in the model than in \snnfd.  We can
improve the ratio of the two \siii\ features by increasing the
luminosity and therefore the temperature of the model, but this would
change the overall shape of the spectrum. The flux levels in the red,
and blue of the \cahk\ feature, would not both match the observation.
The part of the spectrum just blueward of \cahk\ is a very sensitive
temperature/luminosity diagnostic. The photometric colors of the
synthetic spectrum and observations are consistent.

\subsubsection{15 March 1994}

\begin{figure}
\begin{center}
\psfig{file=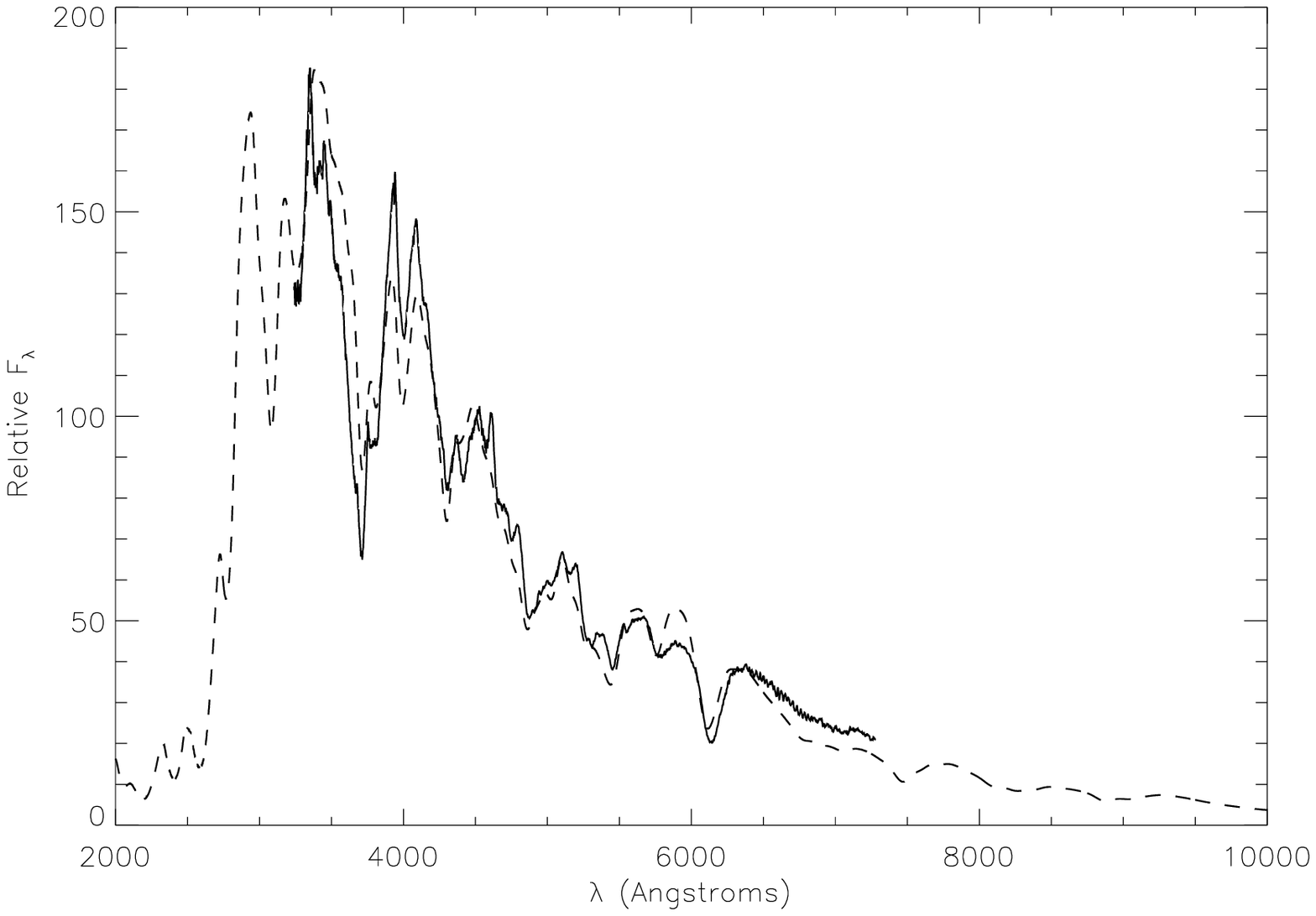,width=14cm}
\caption{\snnfd\ on 15~March 1994 \protect\citep[solid line,][]{filipasi97}
and W7 best fit synthetic spectrum for
day 14 after explosion (dashed line).\label{fig:d14}}
\end{center}
\end{figure}

\begin{figure}
\begin{center}
\psfig{file=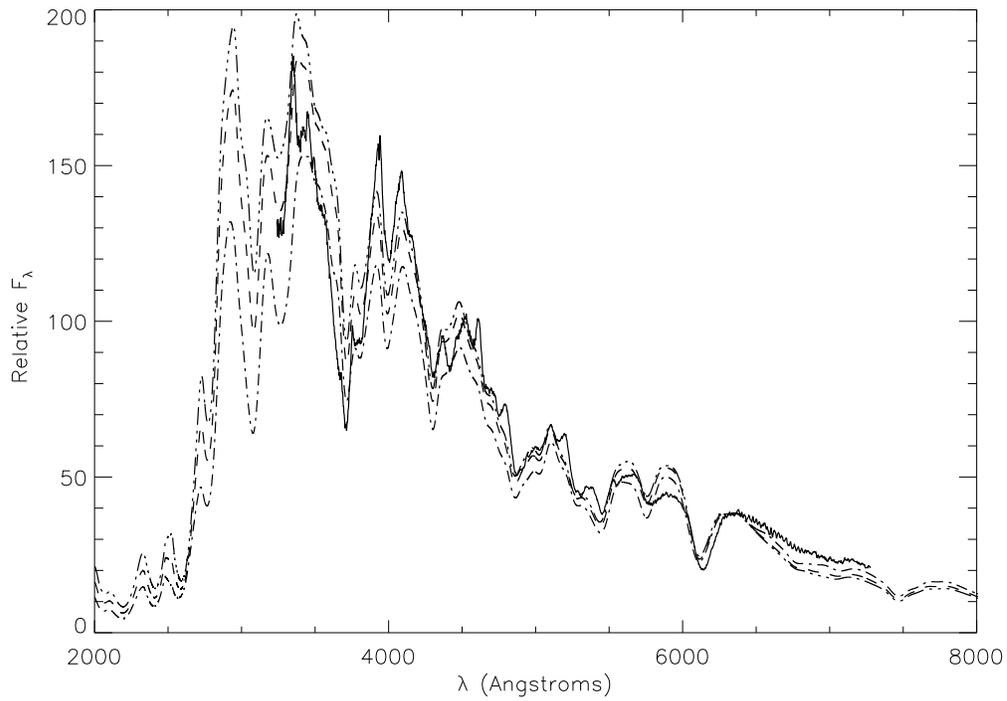,width=14cm}
\caption{\snnfd\ on 15~March 1994 \protect\citep[solid line,][]{filipasi97}, 
W7 best fit (dashed line),
best fit~-7\% luminosity (dot-dashed line), and best fit +7\% luminosity
(dot-dot-dashed line) synthetic spectra for
day 14 after explosion.\label{fig:d14comp}}
\end{center}
\end{figure}

The  \snnfd\ observed spectrum of day~14 (15~March 1994) with our best
synthetic spectrum is shown in Figure~\ref{fig:d14}.  The fit to the
shape and features is good. The colors are in agreement, except \rmi,
which is bluer than the observations. This is the result of a modest
red deficit redward of the \siii\ features. In order to estimate the
accuracy of our parameter determination,
we display the best fit model and observation with the models with
luminosities $\pm 7\%$\ from the best fit in Figure~\ref{fig:d14comp}.
The differences are largest around the \cahk\ feature. This shows how
accurately the fits can be made if spectra are available extending
blueward of the \cahk\ feature and the input model provides a well fit
spectrum.

\subsubsection{16 March 1994}

\begin{figure}
\begin{center}
\psfig{file=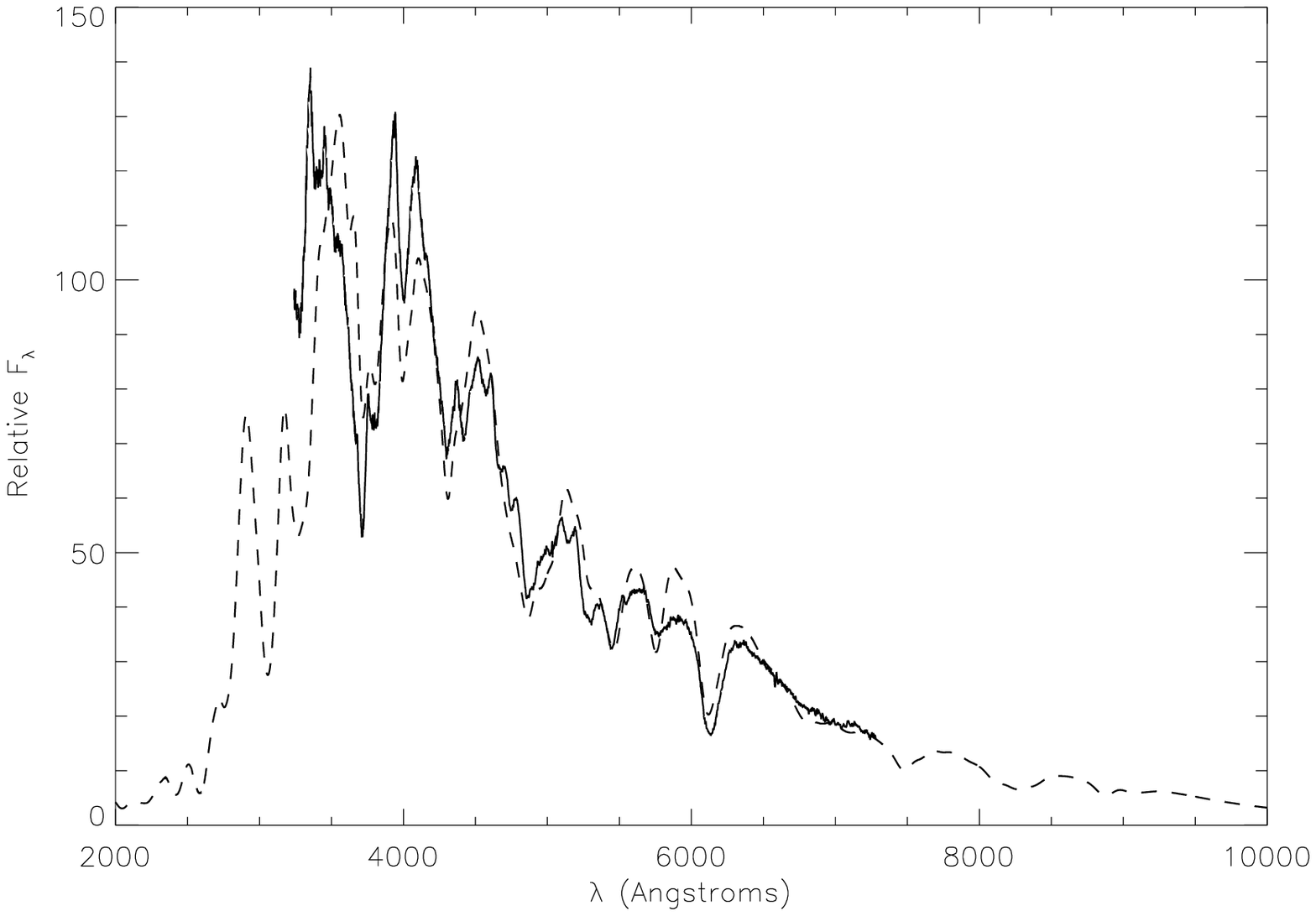,width=14cm}
\caption{\snnfd\ on 16~March 1994 (solid line, Filippenko 1994,
private communication)
and W7 best fit synthetic spectrum for
day 15 after explosion (dashed line).\label{fig:d15}}
\end{center}
\end{figure}

We plot the 16~March 1994 observed spectrum of \snnfd\ with our best synthetic
spectrum at day~15 in Figure~\ref{fig:d15}.
The fit has a good shape and many of the features fit. The \cahk\ feature
does not absorb blueward enough, and the feature just blueward absorbs
much too strongly.
The synthetic colors are good, especially the \umb\ which agrees
very well with \citet{patat94D96}. The \rmi\ color is significantly
bluer than the observations, possibly due to the steeper slope at the red
edge of the \siii\ feature.

\subsubsection{17 March 1994}

\begin{figure}
\begin{center}
\psfig{file=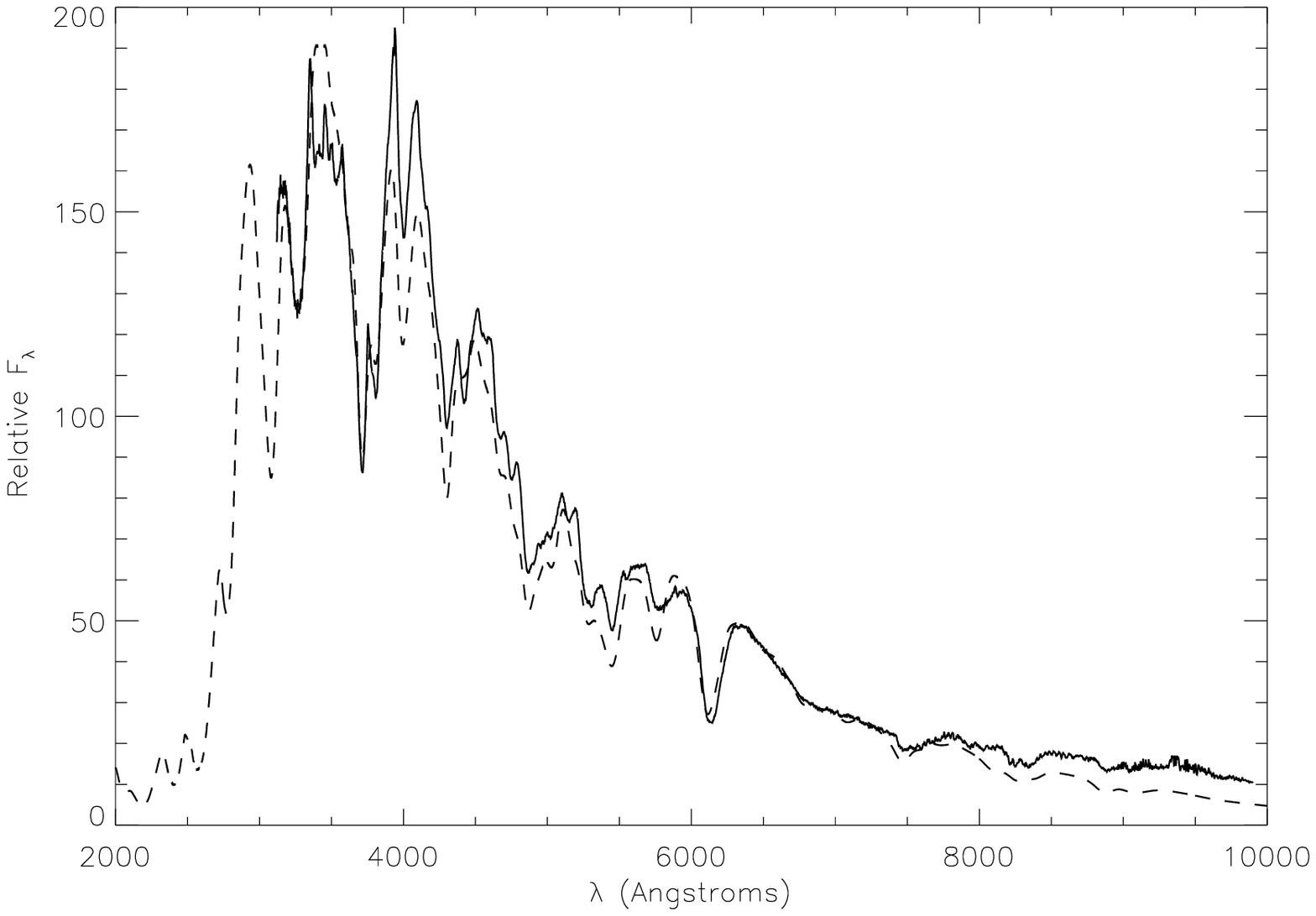,width=14cm}
\caption{\snnfd\ on 17~March 1994 (solid line, Filippenko 1994,
private communication)
and W7 best fit synthetic spectrum for
day 16 after explosion (dashed line).\label{fig:d16}}
\end{center}
\end{figure}

The 17~March 1994 observed spectrum of \snnfd\ with our best synthetic
spectrum at day~16 are shown in Figure~\ref{fig:d16}.  The synthetic
spectrum fits the observation very well. The shapes of the features
match very well. Of note is the fit in the \cahk\ and blueward, and
the fit of the \siii\ feature at 6150~\AA\, including the slope and
`bend' in the continuum redward of the feature. There is only the
small downward continuum shift between these two well fit
regions. This shift is reflected in the \vmr\ color which is 0.10 to
0.15 magnitudes redder depending on the comparison observations. The
rest of the colors coincide with the observations.  For this model the
total continuum optical depth at 5000~\AA, \tstd, is unity at
9000~\kmps. This is the transition point between the partially burned
material and the \nifs-rich core.

\subsubsection{18 March 1994}

\begin{figure}
\begin{center}
\psfig{file=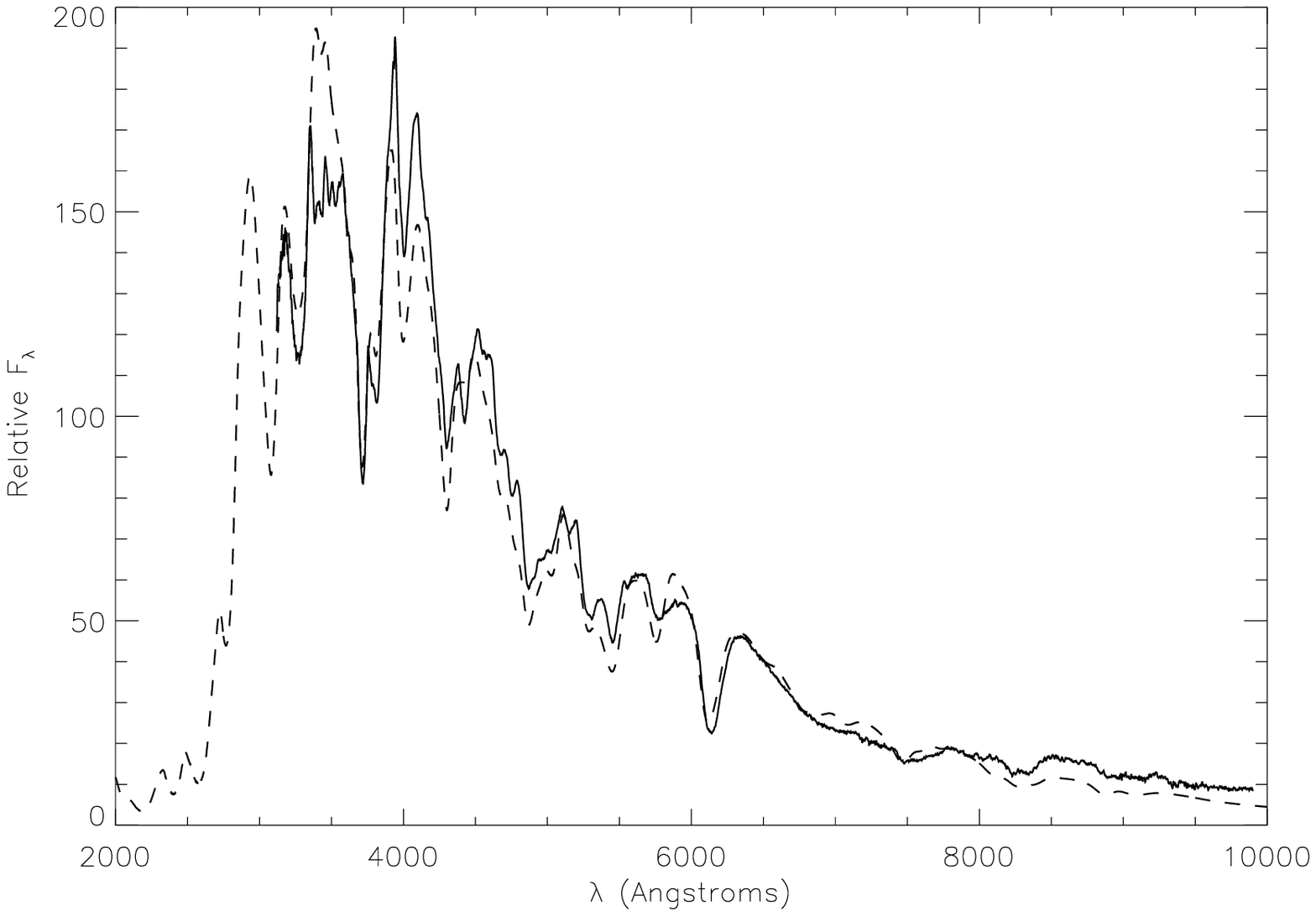,width=14cm}
\caption{\snnfd\ on 18~March 1994 \protect\citep[solid line,][]{filipasi97}
and W7 best fit synthetic spectrum for
day 17 after explosion (dashed line).\label{fig:d17}}
\end{center}
\end{figure}

We plot the 18~March 1994 observed spectrum of \snnfd\ with our best
synthetic spectrum at day~17 in Figure~\ref{fig:d17}.  Like day~16,
the day~17 spectrum fits the \cahk\ feature very well. The remaining
feature shapes are also good, but the slightly lower continuum level
between the \cahk\ and \siii\ features remains. Again, this makes the
\vmr\ color redder than the observations.

\subsubsection{19 March 1994}

\begin{figure}
\begin{center}
\psfig{file=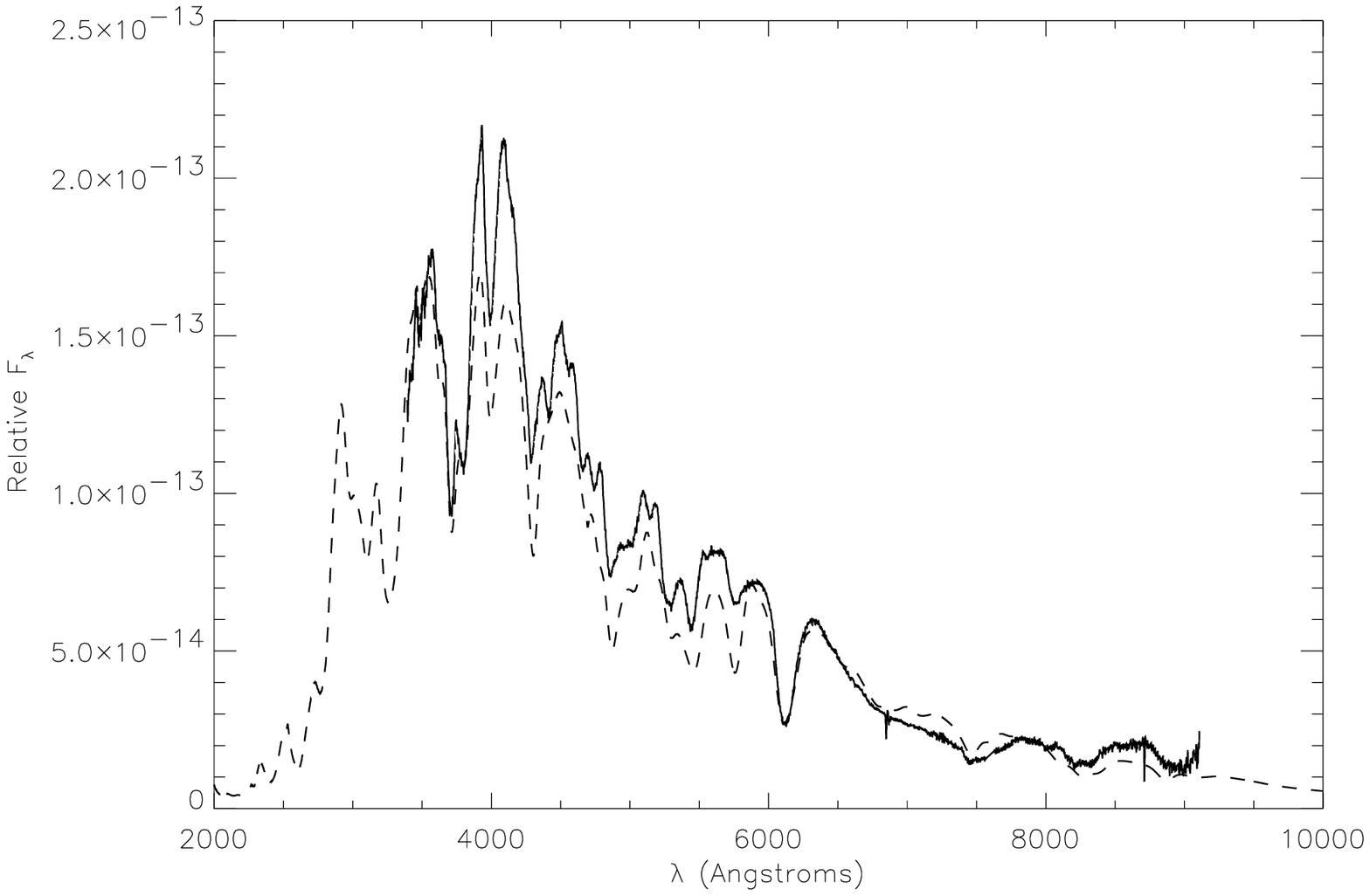,width=14cm}
\caption{\snnfd\ on 19~March 1994 \protect\citep[solid line,][]{patat94D96}
and W7 best fit synthetic spectrum for
day 18 after explosion (dashed line).\label{fig:d18}}
\end{center}
\end{figure}

Figure~\ref{fig:d18} displays the 19~March 1994 observed spectrum of
\snnfd\ with our best 
synthetic spectrum at day~18.  The shape of
the fit is good, especially blueward of 4000~\AA\ and redward of
6000~\AA. The shape of the features between the \cahk\ and 6150~\AA\
\siii\ features is good, but they are shifted to lower flux. This
lower flux makes the \vmr\ color redder than the observations.

\subsubsection{20 March 1994\label{day19}}

\begin{figure}
\begin{center}
\psfig{file=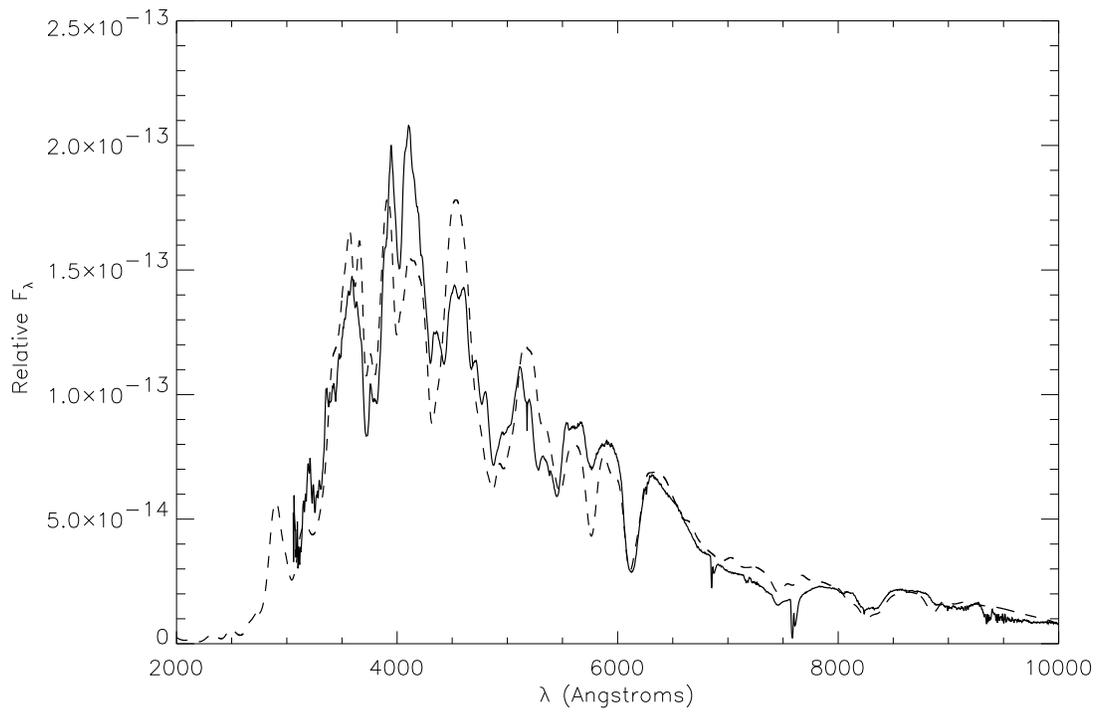,width=14cm}
\caption{\snnfd\ on 20~March 1994 (solid line, data obtained on the
WHT with ISIS by Nic Walton and reduced at the RGO by Jim Lewis, 
private communication)
and W7 best fit synthetic spectrum for
day 19 after explosion (dashed line).\label{fig:d19}}
\end{center}
\end{figure}

Plotted in Figure~\ref{fig:d19} is the 20~March 1994 observed spectrum
of \snnfd\ with our best synthetic spectrum at day~19.  The shape of
the synthetic spectrum is good. The fit is good blueward of 4000~\AA\
and redward of 6000~\AA.  The shapes of many of the features between
are poor. The overly strong blue \siii\ feature at 5750~\AA\ indicates
that the temperature of the line-forming region is too low. By
increasing the luminosity by 40\%, the resulting temperature change
improves the fit of the blue \siii\ feature and the \siiw\ feature at
5000-5500~\AA. This badly changes the shape of the blue part of the
synthetic spectrum. These features form in the optically thin zone at
this and later epochs. If we could heat this zone without affecting
the luminosity the quality of the spectrum could be improved.  A
viable way to achieve this would be to mix additional \nifs\ into the
zone, which is expected in multi-dimensional burning models.  The
colors are all redder than the observations, particularly \bmv\ and
\rmi, thus suggesting that the luminosity or model temperature is too
low.

\subsection{After Visual Maximum\label{postmax}}

\subsubsection{23 March 1994}

\begin{figure}
\begin{center}
\psfig{file=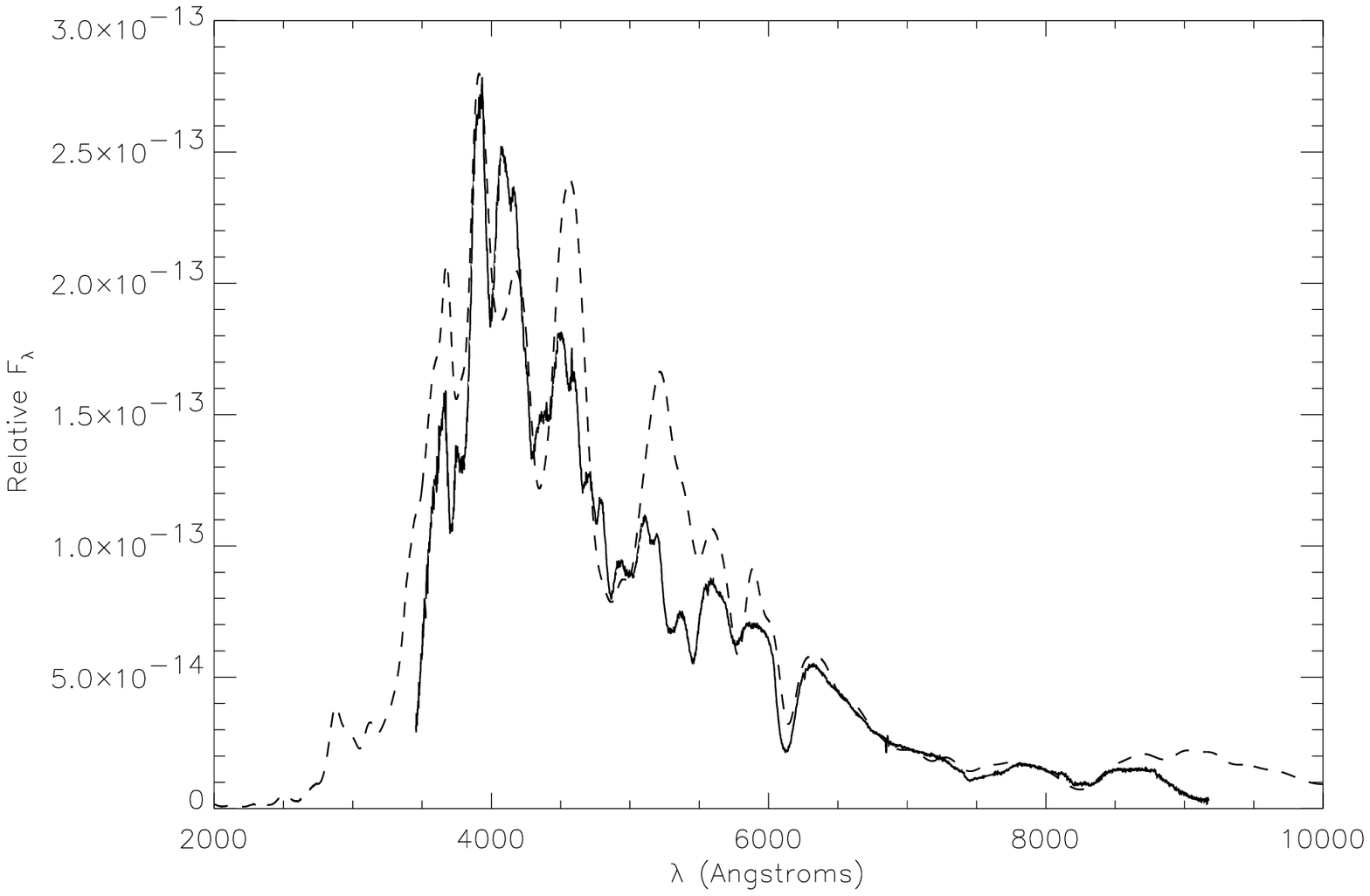,width=14cm}
\caption{\snnfd\ on 23~March 1994 \protect\citep[solid line,][]{patat94D96}
and W7 best fit synthetic spectrum for
day 22 after explosion (dashed line).\label{fig:d22}}
\end{center}
\end{figure}

We plot the 23~March 1994 observed spectrum of \snnfd\ with our best
synthetic spectrum at day~22 in Figure~\ref{fig:d22}.  The shape of
this spectrum is good, but the strength of some of the optical
features is exaggerated. The \siiw\ is far too weak in our synthetic
spectrum. The observation shows a sharp edged feature which indicates
that the opacity strongly cuts off at higher velocities. This is
consistent with the sulfur distribution in W7. In
the region of strong sulfur concentration, the sulfur in the W7 model
has an ionization ratio 
S$^+/$S$^{++} \sim 2$.  This, along with the temperature sensitive
strength of the red \siii\ feature at 5800~\AA, suggests that the
ionization temperature should be higher in the zone between the
Fe-core and the unburned region in these epochs where they are
optically thin. The colors are consistent with the trends in the
observations, except the \vmr, which is affected by the larger fluxes
at the \siiw\ feature.

\subsubsection{26 March 1994}

\begin{figure}
\begin{center}
\psfig{file=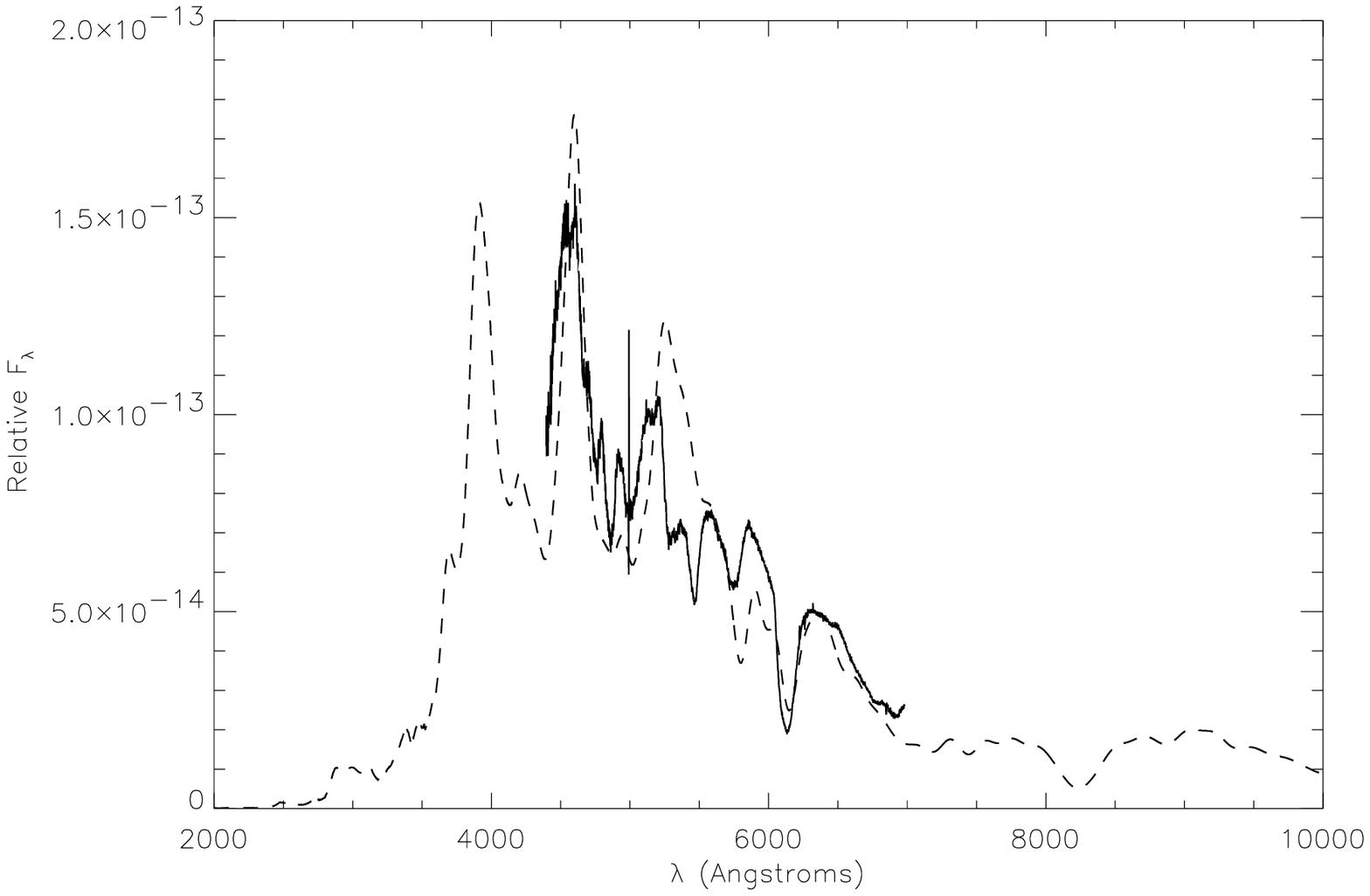,width=14cm}
\caption{\snnfd\ on 26~March 1994 \protect\citep[solid line,][]{patat94D96}
and W7 best fit synthetic spectrum for
day 25 after explosion (dashed line).\label{fig:d25}}
\end{center}
\end{figure}

We plot the 26~March 1994 observed spectrum of \snnfd\ with our best
synthetic spectrum at day~25 in Figure~\ref{fig:d25}.  As in day~22,
the \siiw\ is weak and the red \siii\ feature is too strong.  The
overall shape of the spectrum appears to be good, but the limited
range of the spectral data make a more certain fitting of the blue
peaks near \cahk, and therefore the total luminosity, difficult. This
is reflected in the colors, which are erratic, and do not follow any
general trend. Of particular note is the shift in the synthetic \umb\
color to redder than the observations, indicating a near-$UV$ deficit in
the synthetic spectra.

\subsubsection{28 March 1994}

\begin{figure}
\begin{center}
\psfig{file=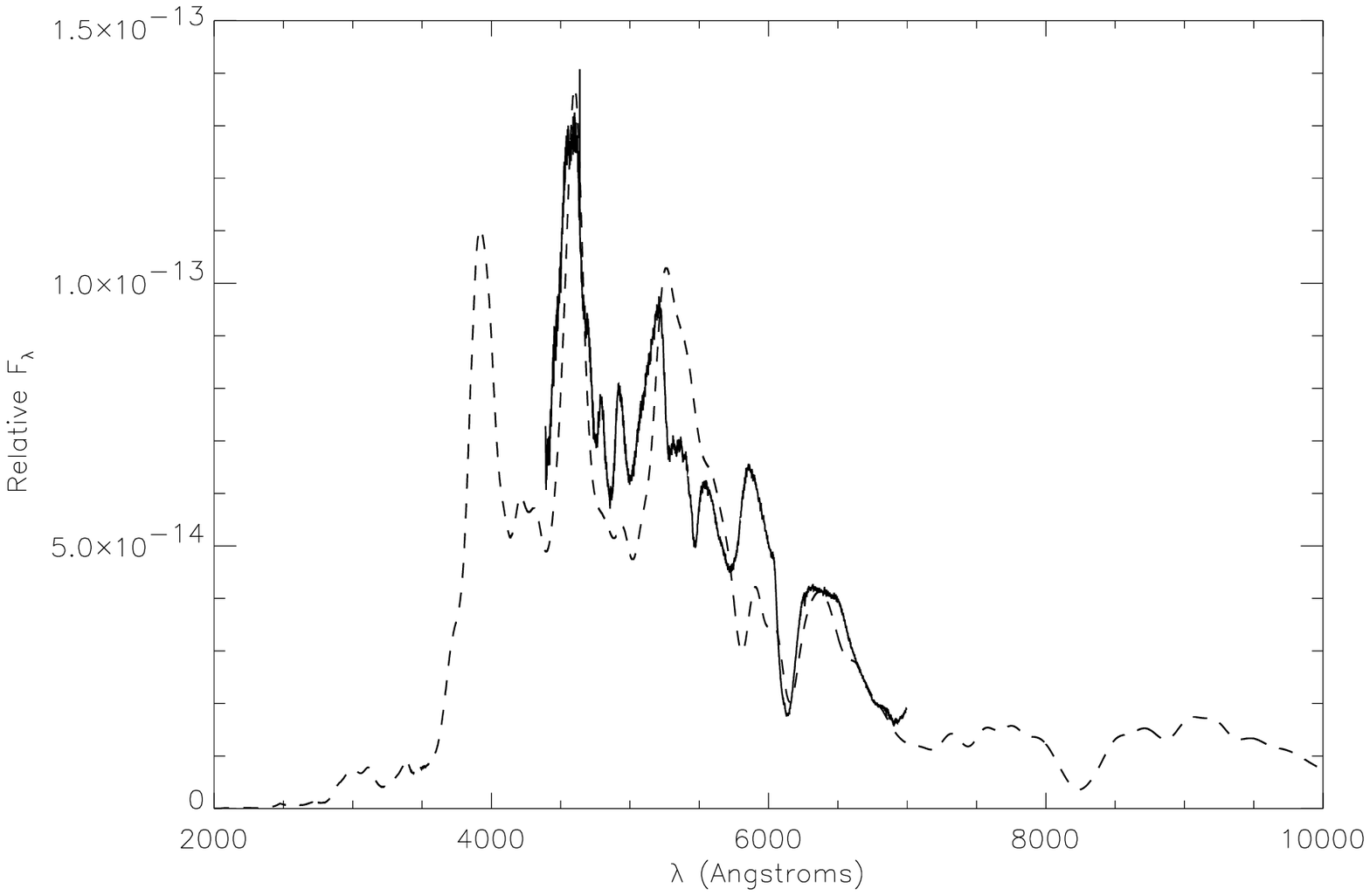,width=14cm}
\caption{\snnfd\ on 28~March 1994 \protect\citep[solid line,][]{patat94D96}
and W7 best fit synthetic spectrum for
day 27 after explosion (dashed line).\label{fig:d27}}
\end{center}
\end{figure}

Plotted in Figure~\ref{fig:d27} is the 29~March 1994 observed spectrum
of \snnfd\ with our best synthetic spectrum at day~27.  As on day~25
the limited range of the available observed spectrum makes precise
fitting of the luminosity difficult. The general difficulties in
making the rest of the spectrum fit may be due to problems with the W7
model or with the temperature structure we have calculated in our
quasi-static, energy balanced models. The spectrum is also making the
nebular transition, the continuum optical depth at 5000~\AA, $\tstd
\sim 4$. Like day~25, the colors disagree with the observed
photometry.

\subsubsection{31 March 1994}

\begin{figure}
\begin{center}
\psfig{file=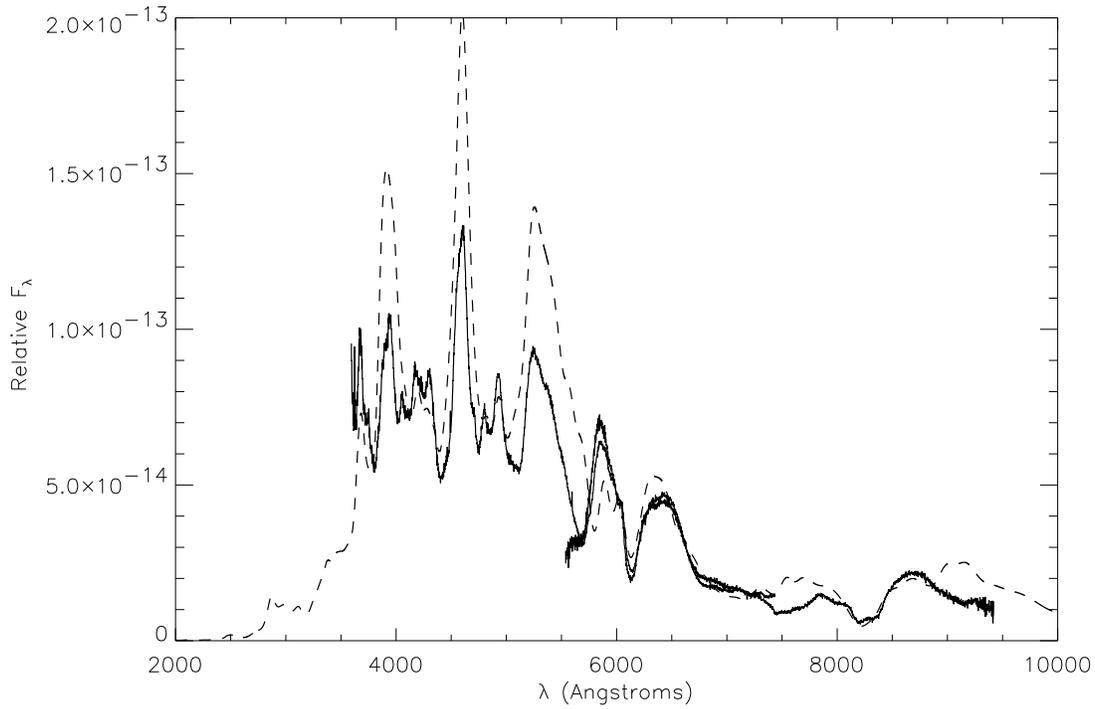,width=14cm}
\caption{\snnfd\ on 31~March 1994 
\protect\citep[solid line, red portion,][]{patat94D96} and W7 best fit
synthetic spectrum for day 30 after explosion (dashed line). The
scaled spectrum 
from 2~April 1994 
\protect\citep[solid line, blue portion,][]{patat94D96}
has been added to aid in
fitting.\label{fig:d30}}
\end{center}
\end{figure}

We plot the 31~March 1994 observed spectrum of \snnfd\ with our best
synthetic spectrum at day~30 in Figure~\ref{fig:d30}.  The minimum
wavelength of the 31~March spectrum is only 5500~\AA, so we have
scaled the 2~April spectrum and added it to Figure~\ref{fig:d30}. We
have now switched to nebular boundary conditions because $\tstd \sim
3$ [see \citet{nugphd} and \citet{nughydro97} for a discussion of our
choice of boundary conditions and the results of tests].
This means that we assume that the space inside the innermost zone, at
$\sim 1000$~\kmps, is completely transparent. The shape of the
spectrum is good and the line features are acceptable. The largest
discrepancy is the complete lack of the \naid\ feature.  This feature
has been slowly growing in the observations, but is not replicated in
the synthetic spectra. We discuss this in \S~\ref{cana}. The colors
are still erratic, but the \umb\ is not as excessively red as before.

\subsubsection{2 April 1994}

\begin{figure}
\begin{center}
\psfig{file=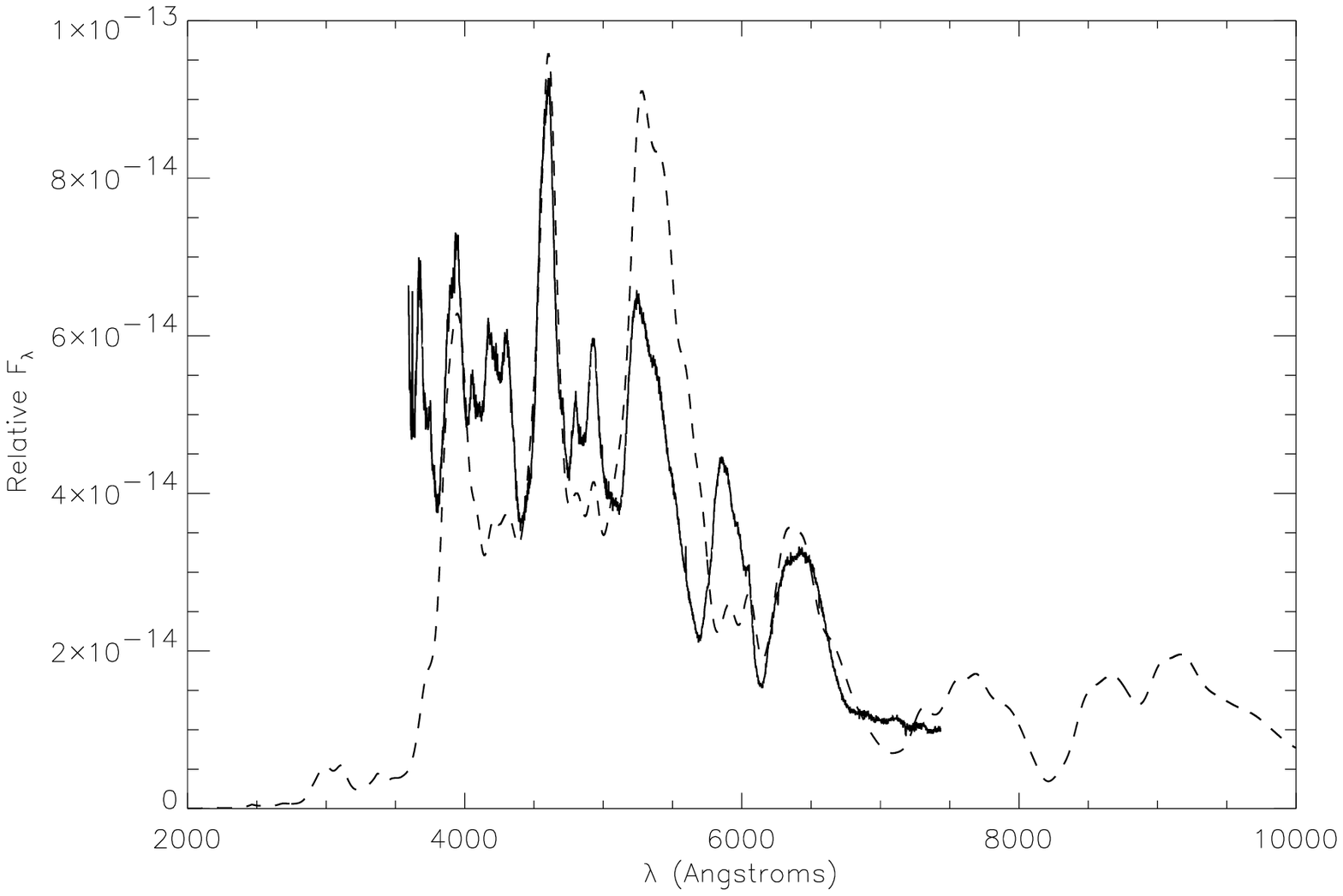,width=14cm}
\caption{\snnfd\ on 2~April 1994 
\protect\citep[solid line,][]{patat94D96} 
and W7 best fit synthetic spectrum for
day 32 after explosion (dashed line).\label{fig:d32}}
\end{center}
\end{figure}

Figure~\ref{fig:d32} shows the 2~April 1994 observed spectrum of
\snnfd\ with our best synthetic spectrum at day~32.  This spectrum
fits a bit less well than for day~30. Again the strong \naid\ feature
is missing. Like the day~30 model, the day~32 model uses nebular
boundary conditions and the luminosity corresponds to instantaneous
re-processing of deposited \gamray\ energy. The continuum optical
depth continues to decline and for this model is $\tstd \sim 2$. The
colors continue to be erratic, and the deficit in the near-$UV$ is
reflected by the very red \umb.

\subsection{Pre-observation Spectra} 

\begin{figure}
\begin{center}
\psfig{file=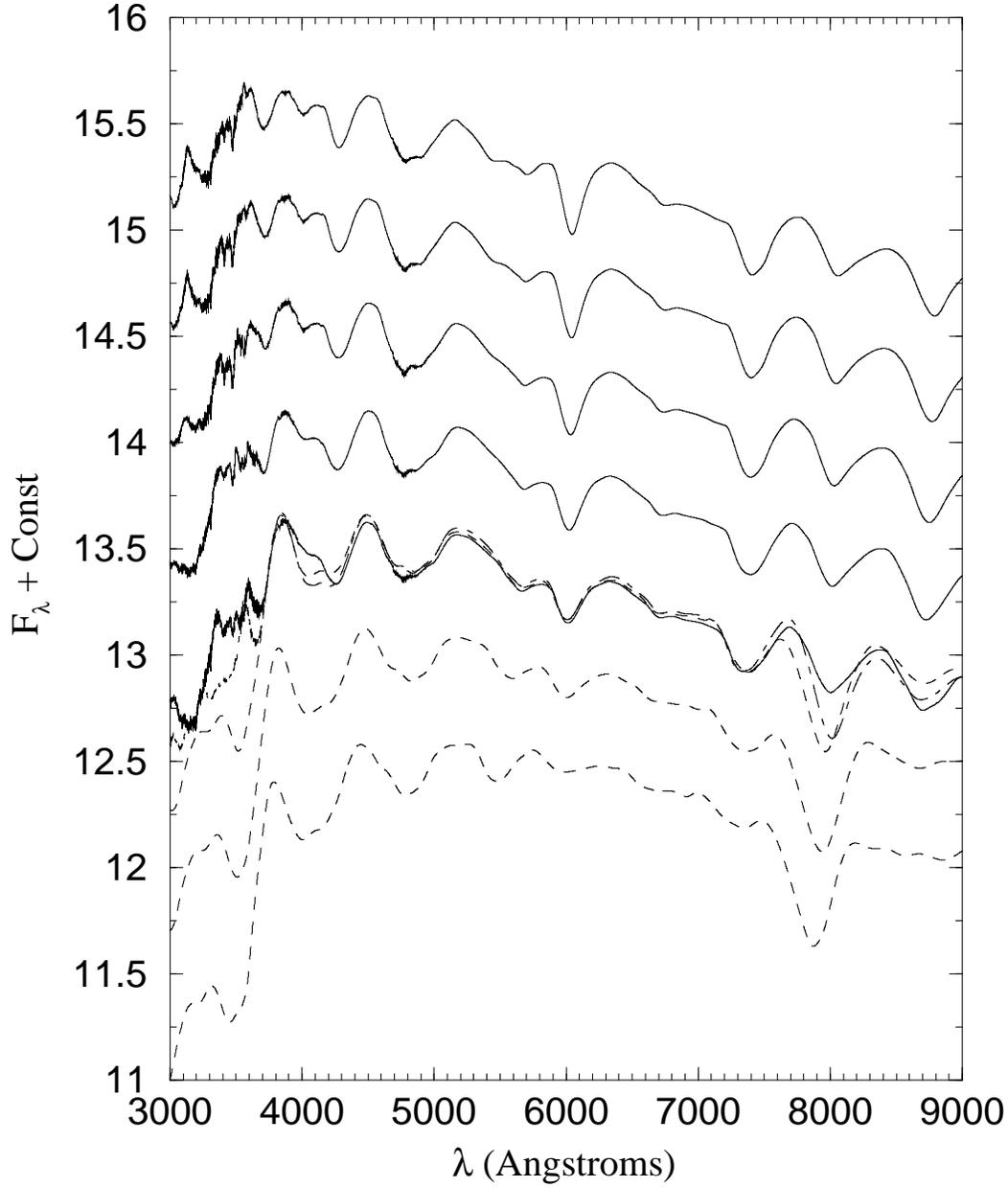,width=14cm,angle=270}
\caption{Synthetic spectra for the projected luminosities for
days 1--7 (from bottom to top) after explosion. The dashed lines are
the exponential models, and the dot-dashed line in day 3 is the
exponential model that has been truncated at $v_\textrm{max} =
30000$~\kmps.\label{fig:early}} 
\end{center}
\end{figure}

We have extended the luminosities found by fitting the earliest
spectra using a parabolic luminosity law, $L \propto t^2$, as used by
\citet{riessetalIaev00} and \citet{akn00}. This form approximates the
early \snia\ as a fixed temperature object with a constant
\emph{velocity} photosphere.  We plot our synthetic spectra using our
extended luminosities in Figure~\ref{fig:early}.  To avoid
computational problems caused by the sharp density `spikes' at the
edge of the unburned C+O which are close to the photosphere for the
day~1 and day~2 models, we have used an exponential density law, $\rho
\propto e^{-v/v_e}$, where $v_e = 2700$~\kmps\ to match the shape of
the W7 density profile.  The three exponential models (dashed lines)
extend to $\sim 60000$~\kmps\ and show a much larger $UV$-deficit at the
\cahk\ edge. We have computed another model at day~3 where the outer
edge of the ejecta was at $\sim 30000$~\kmps, as it is in W7,
demonstrating that the strength of the large $UV$ deficit is related to
the extension of the atmosphere. This will be an interesting topic for
future modeling of early \sneia\ $UV$ spectra and photometry.  The other
significant difference between the exponential and regular W7 models
is the \cair\ feature at 8000~\AA.

The features in the synthetic spectra evolve during the week after
explosion.  The \siii\ feature becomes weaker for earlier
spectra. This behavior is the same as found in \citet{lentzmet00},
where some features, including \siii, were formed mostly in the
intermediate velocity regions containing freshly formed silicon. At
early times, the outer layers are more opaque to the deeper,
silicon-rich material and the \siii\ feature forms in the unenriched
C+O layers. Additionally at earlier times, the mass of silicon outside
the ``photosphere'', loosely defined as $\tstd \sim 1$, is less than
later times. This decrease in absorption is also likely the reason for the
decreasing strength of the \ion{O}{1}\ feature at 7500~\AA. The \feii\
feature at 4800~\AA\ is formed in the fast C+O layer as we have
discussed in \S~\ref{mar9}. The \cair\ is likely saturated and only
changes with the switch to an exponential density profile.

\section{Synthetic Photometry}

\begin{figure}
\begin{center}
\psfig{file=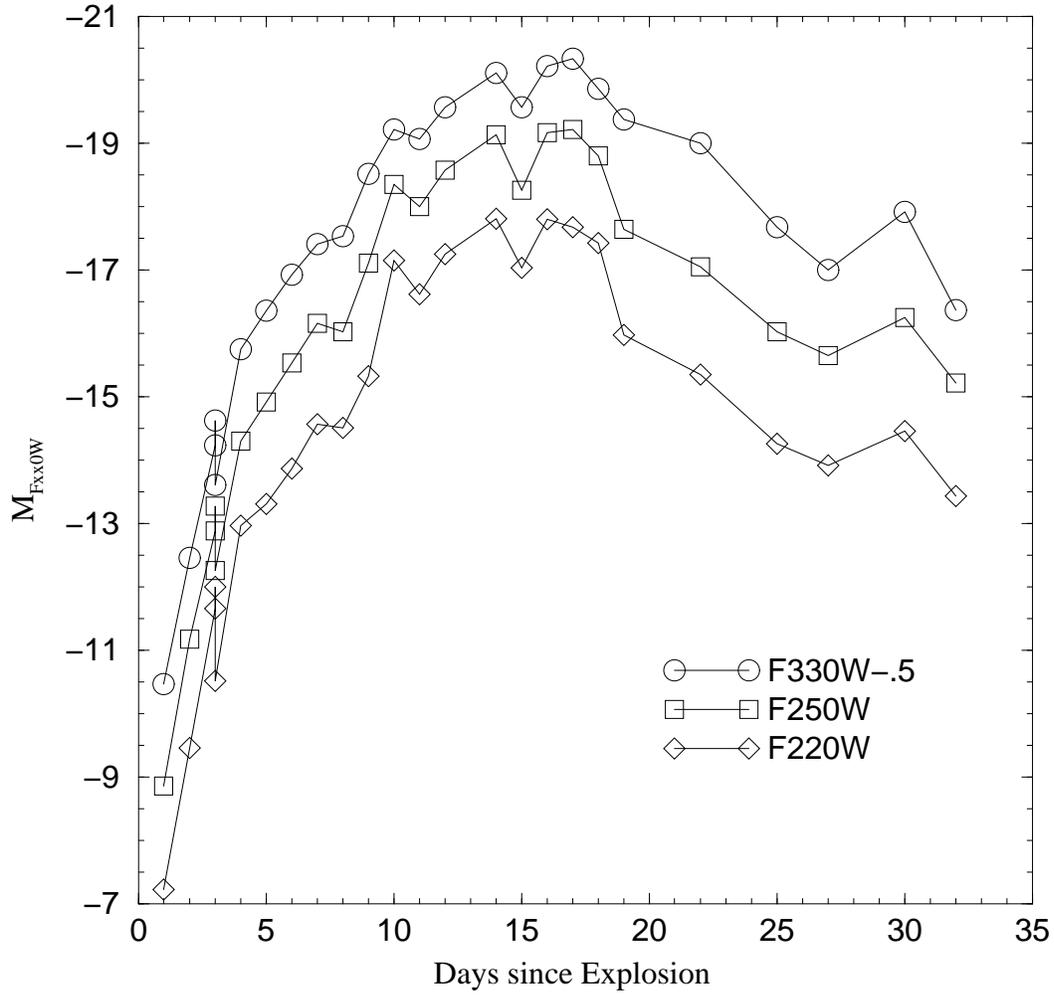,width=14cm,angle=270}
\caption{Ultra-violet synthetic photometry from the synthetic spectra.
Arbitrary shifts in magnitude are noted in the legend in this and the
following figures\label{fig:photuv}.}
\end{center}
\end{figure}

\begin{figure}
\begin{center}
\psfig{file=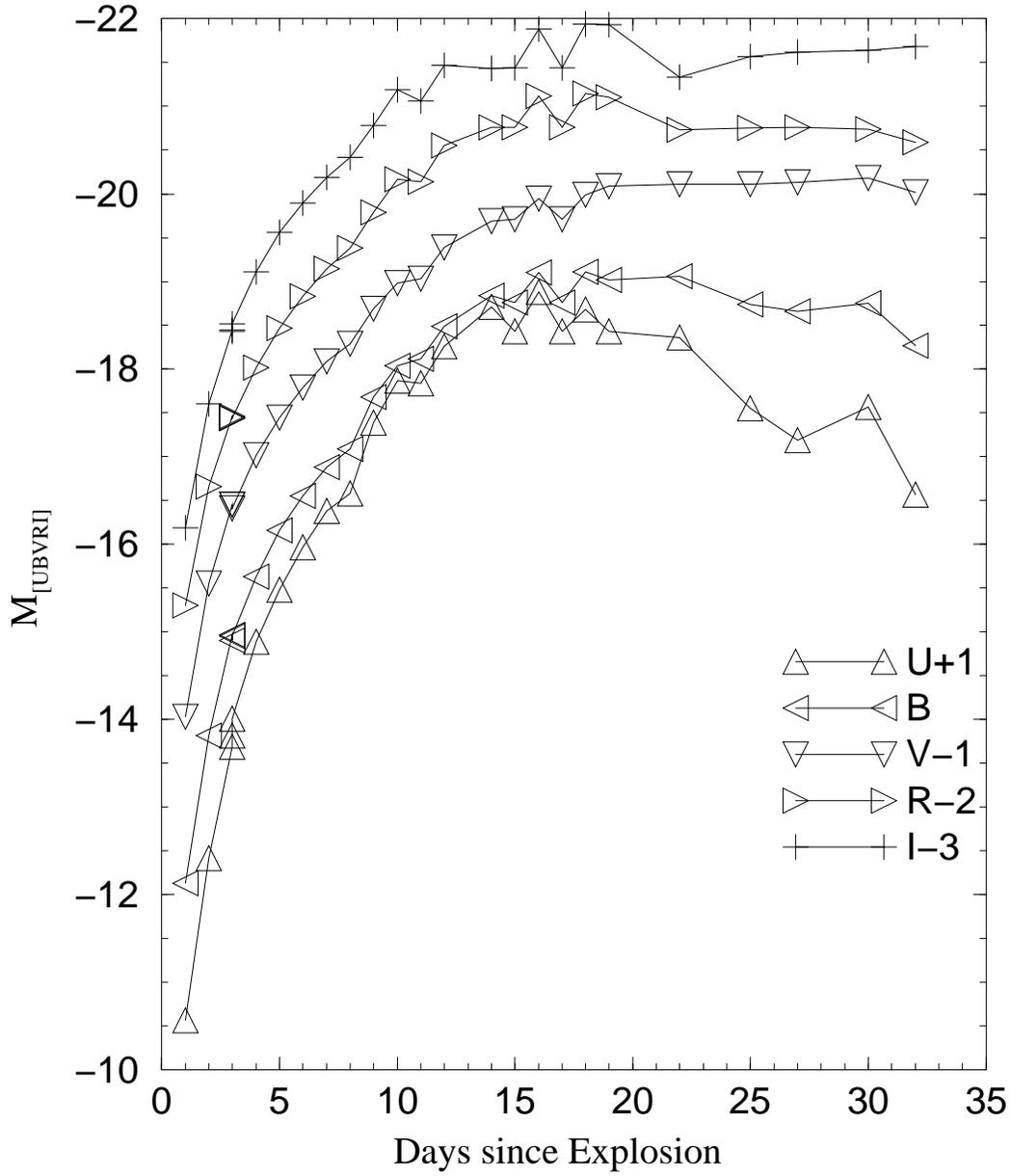,width=14cm,angle=270}
\caption{Synthetic optical/near-$IR$ photometry from the synthetic spectra.
\label{fig:photopt}}
\end{center}
\end{figure}

\begin{figure}
\begin{center}
\psfig{file=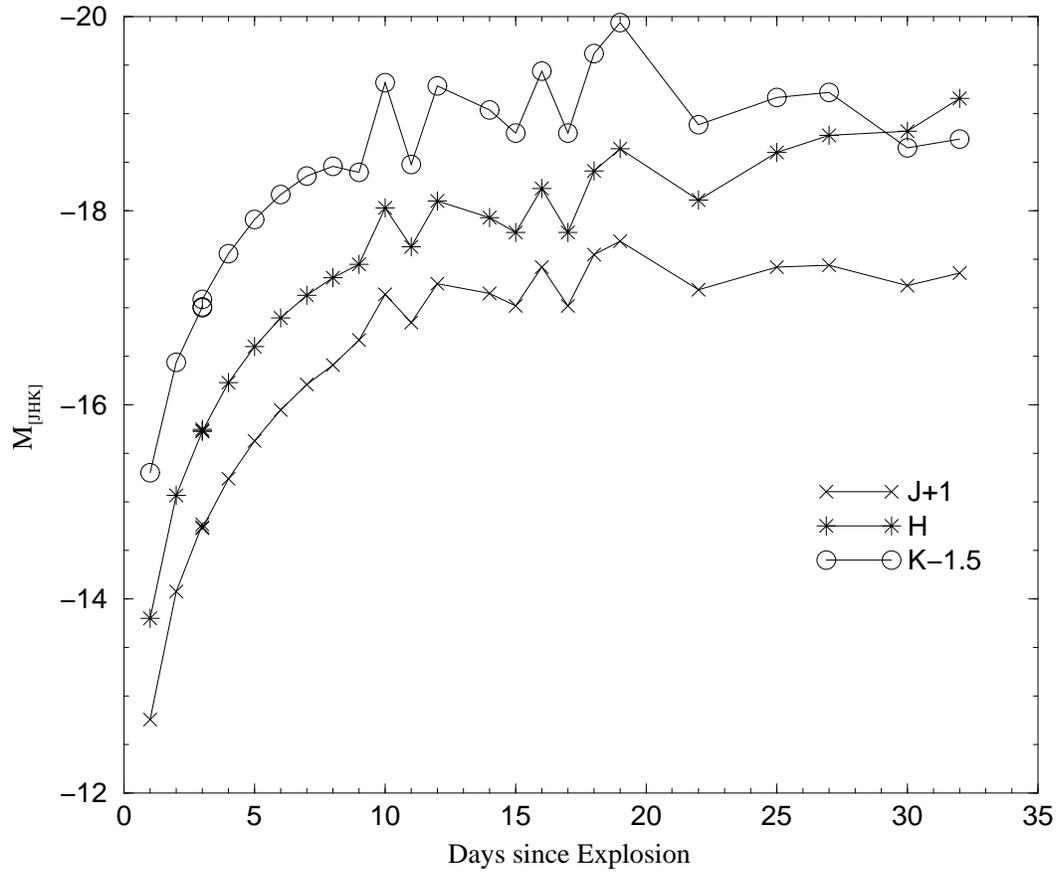,width=14cm,angle=270}
\caption{Infra-red synthetic photometry from the synthetic spectra.
\label{fig:photir}}
\end{center}
\end{figure}

\begin{figure}
\begin{center}
\psfig{file=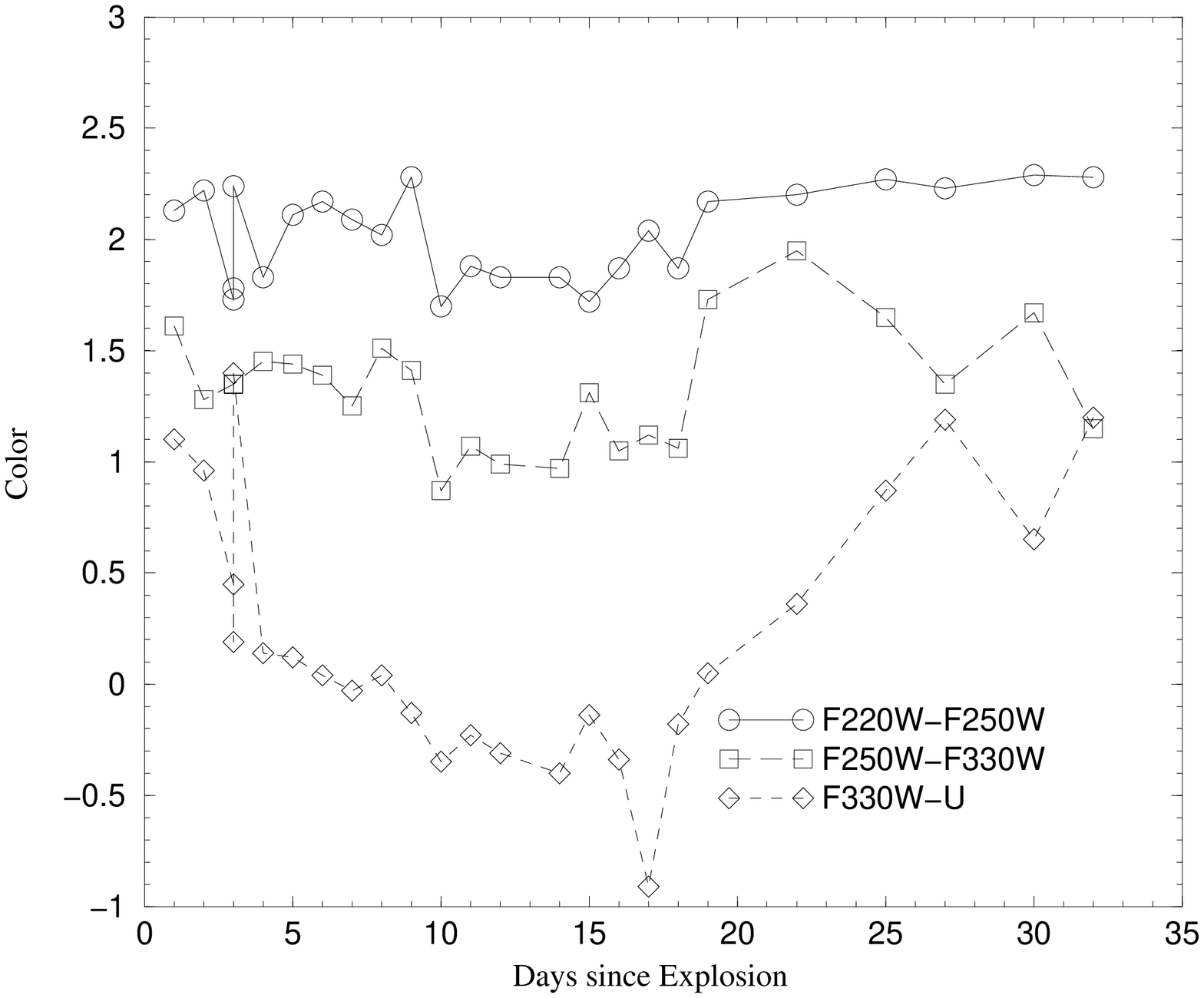,width=14cm,angle=270}
\caption{Ultra-violet synthetic colors from the synthetic spectra.
\label{fig:coloruv}}
\end{center}
\end{figure}

\begin{figure}
\begin{center}
\psfig{file=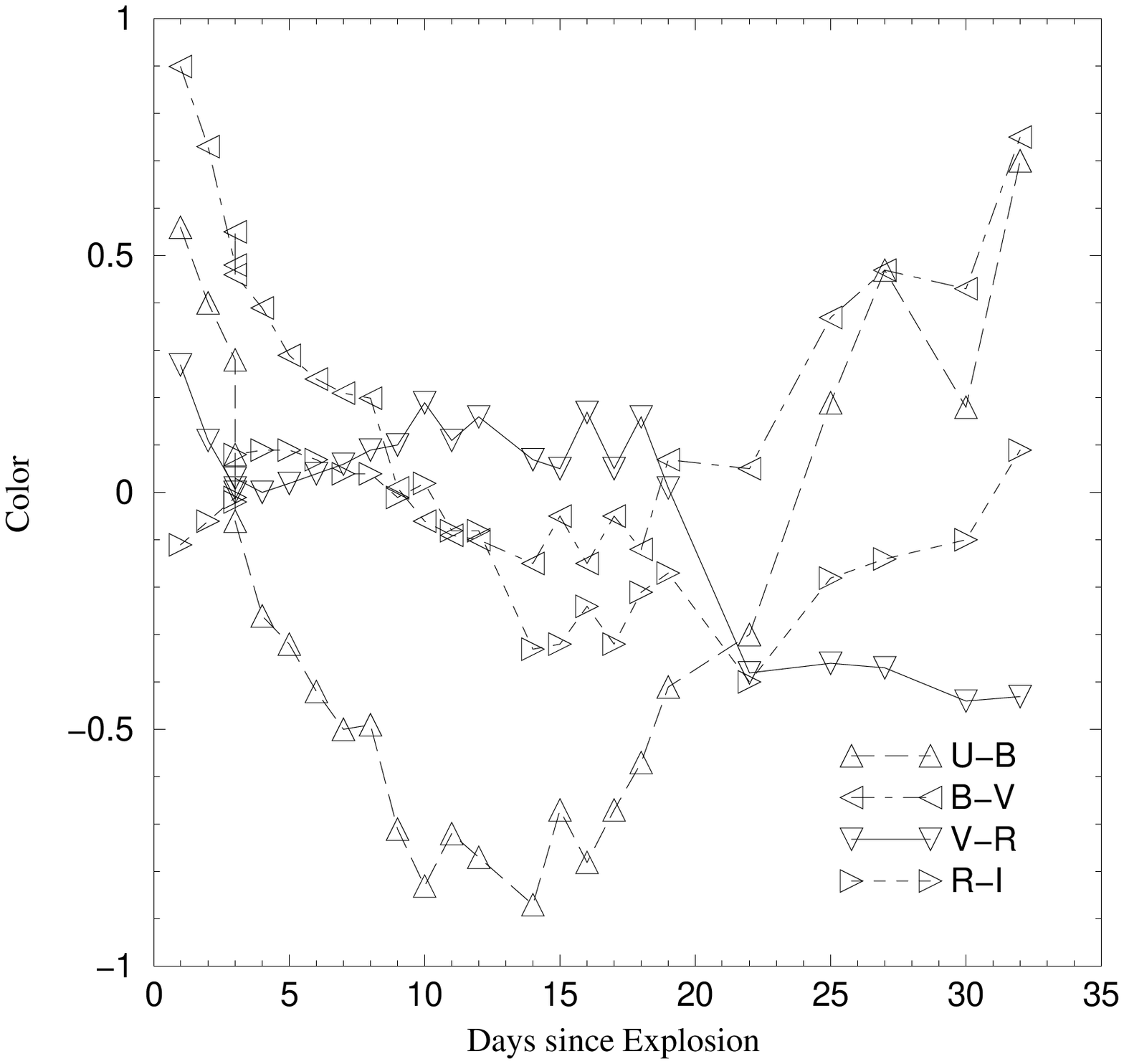,width=14cm,angle=270}
\caption{Synthetic colors from the synthetic spectra.
\label{fig:coloropt}}
\end{center}
\end{figure}

\begin{figure}
\begin{center}
\psfig{file=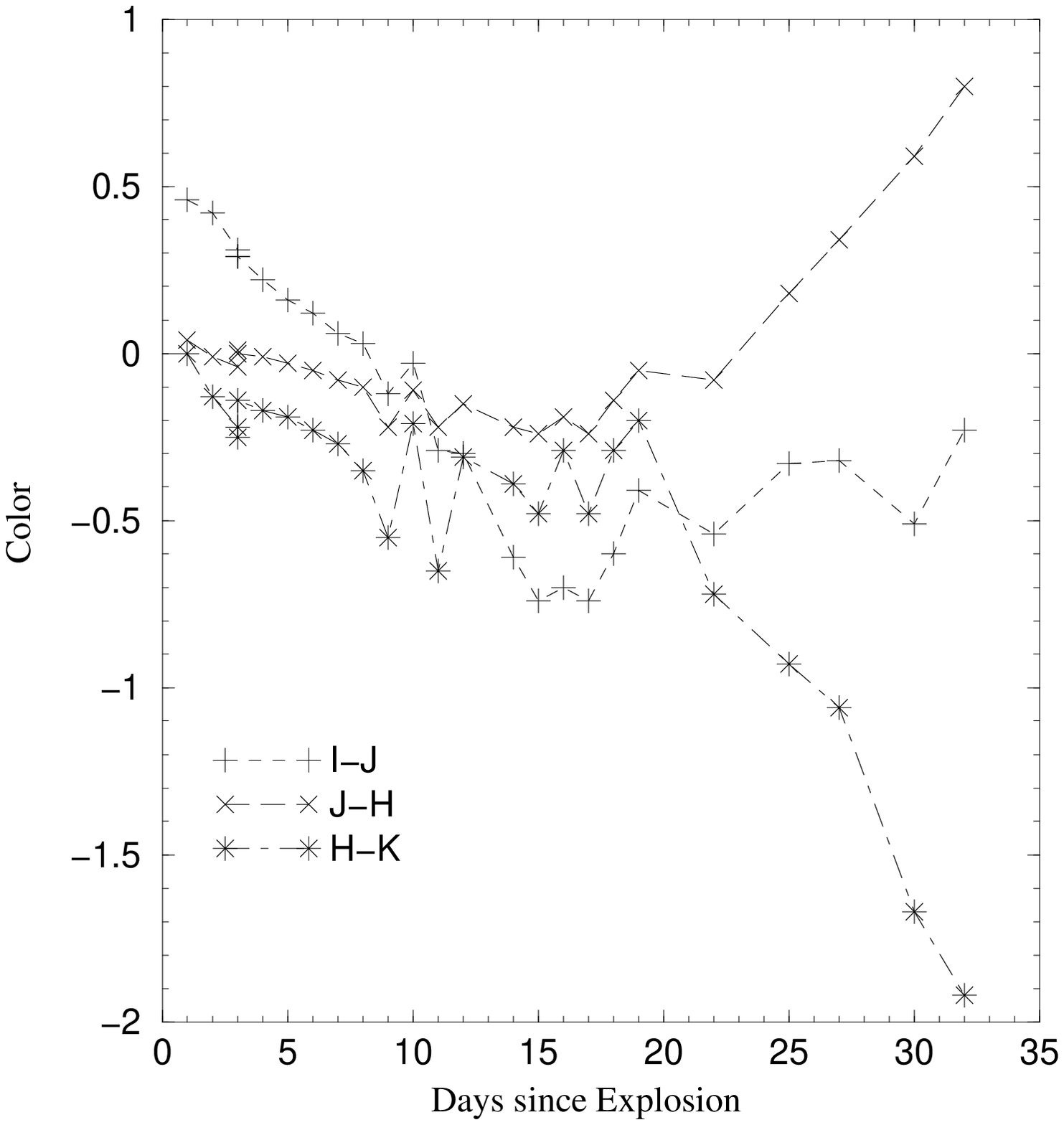,width=14cm,angle=270}
\caption{Infra-red synthetic colors from the synthetic spectra.
\label{fig:colorir}}
\end{center}
\end{figure}

We have computed synthetic $UBVRI$ \citep{bessell90} using the
prescription of \citet{hamuyetal92}, and $JHK$ (M.~Hamuy,
private communication) photometry from the synthetic spectra presented
in \S~\ref{spec}.  The synthetic photometry are plotted in
Figures~\ref{fig:photuv}, \ref{fig:photopt}, \&~\ref{fig:photir}, and
the colors in Figures~\ref{fig:coloruv}, \ref{fig:coloropt},
\&~\ref{fig:colorir}.
 
\subsection{Ultra-Violet Photometry and Colors}

We have calculated the photometry for our synthetic spectra in the $UV$
using the Advanced Camera for Surveys (ACS), scheduled to be installed
in the Hubble Space Telescope in 2001. We used the ACS Exposure Time
Calculator\footnotemark[1] to obtain count rates in each filter and
calibrated the results with a Vega calibration spectrum at $V=15$. The
ACS absolute photometry and the count rates for our spectra when
observed at $7.5 \times 10^{25}$~cm, $\mu=31.93$, are reported in
Table~\ref{tab:uv}. We can see that with this instrument, accurate $UV$
photometry of \sneia\ which are a few days old are possible with short
exposures.

\footnotetext[1]{\tt{http://garnet.stsci.edu/ACS/ETC/acs\_img\_etc.html}}

 The $UV$ light curves in Figure~\ref{fig:photuv} peak 3--6~days before
maximum light in the visual band. This is approximately consistent
with the $U$-band peak 
displacements of -1~day \citep{richsn94d} and -2~days
\citep{patat94D96} in the \snnfd\ observations and -1~day in the light
curve analysis of \citet{vacca94d96}.  The peak timing of the $UV$ light
curves is also consistent with the report of \citet{kir92a}.

The colors (Figure~\ref{fig:coloruv}) between the three ACS
filters are fairly flat with no 
consistent trend, but with some oscillations as the $UV$ region is the
most sensitive to small fitting errors in modeling the synthetic
spectra. The F330W-$U$ color is the difference between the `ACS~$U$' and
the \citet{bessell90}~$U$. Though the filters are intended to be
equivalent, the large, strong, features of \sneia\ make simple
transformations difficult.  These difficulties can
be overcome by applying the relevant transformations for the object's
reddening and red-shift, and for the observational filter and
detector, to the synthetic spectra when generating synthetic
photometry for comparison with observations and illustrate the need
for synthetic spectra when comparing photometric observations.

\subsection{Optical Light Curves and Colors}

Figure~\ref{fig:photopt} shows that the $U$-band, like the F330W band,
peaks about 3--6~days before maximum 
light consistent with the observations. The $B$, $V$, $R$, and $I$ light
curves do not fall as fast after maximum as the observations
\citep{patat94D96,richsn94d,meikle94d91t}.  This is likely due to the
large features of the post-maximum spectra that did not fit as well as
the pre-maximum spectra.

The colors are plotted in Figure~\ref{fig:coloropt}. The \umb\ curve
corresponds well with the observations \citep{patat94D96}, getting
bluer until $\umb < -0.5$ at about -5~days, then becoming redder until
$\umb \sim 0$. The final few models become too \umb\ red, but this is
clearly seen in the $UV$-deficits in the related synthetic spectra. The
\bmv\ curves show the same shallow blue minimum near maximum light and
slow reddening afterward as in the observations
\citep{richsn94d,patat94D96}. The \vmr\ curves follow the same blue
evolution from $\vmr \sim 0.2$ at -10~days to $\vmr \sim -0.3$ at
+10~days as observed \citep{richsn94d,patat94D96}. The \rmi\ color
falls from $\rmi \sim 0$ at -10~days to $\rmi \sim -0.4$ and then
returns to $\rmi \sim 0$ afterwards as do the observations
\citep{richsn94d}, but the timing of the synthetic colors is ambiguous
and possibly earlier than the observations.

\subsection{IR Photometry and Colors}

We have plotted the synthetic infra-red light curves in
Figure~\ref{fig:photir}.  The $K$-band photometry seems to be consistent
with the light curve templates of \citet{elias85}. The $J$-band is flat
rather than falling during the post-maximum period we model, and the
$H$-band rises rather than remaining flat.  This is likely due to the
difficulties noted in \S~\ref{postmax} with features not fitting well.
The pre-maximum $IR$ photometry is flat within fitting scatter for all
bands in the 10 days before maximum consistent with the $IR$ photometry
of SN~1998bu \citep{hern98bu00}. The $IR$ colors are presented in
Figure~\ref{fig:colorir}.

\begin{figure}
\begin{center}
\psfig{file=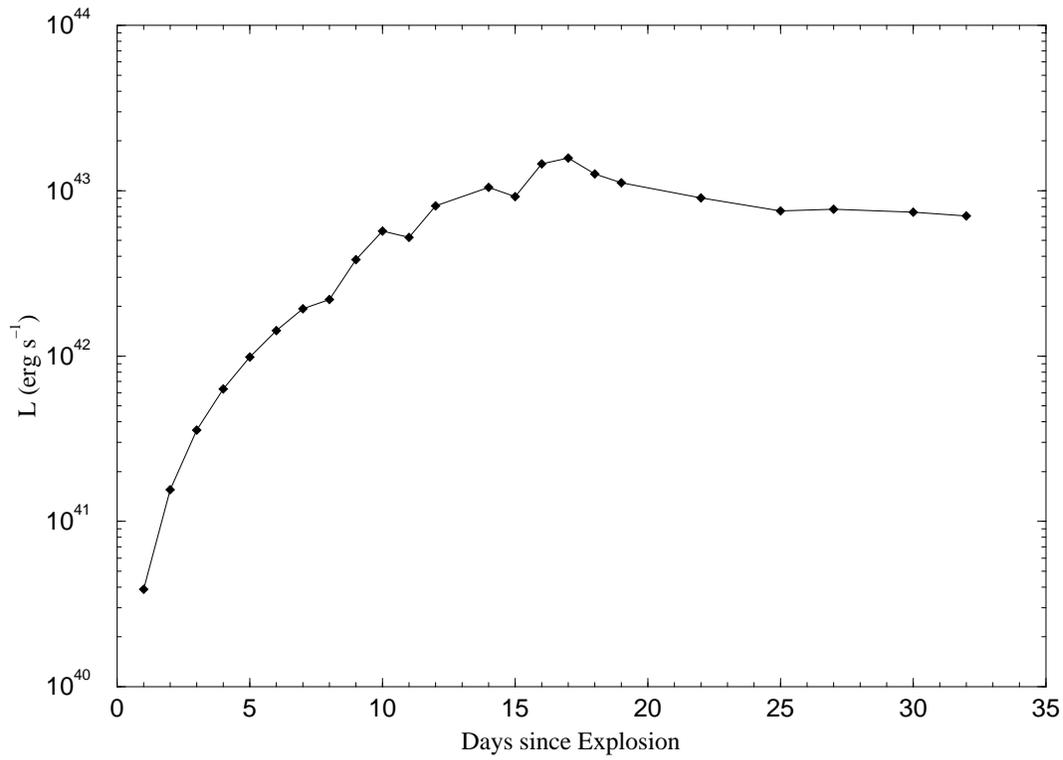,width=14cm,angle=270}
\caption{Bolometric light curve of synthetic
spectra\label{fig:bolo}. The flatness of the bolometric light curve is
likely due to the fact that the luminosity of the peak of W7 is too
low because there was not enough mixing of \nifs, which would force
our fits 
to a larger luminosity at maximum (see \S~\ref{day19}).}
\end{center}
\end{figure}

Figure~\ref{fig:bolo} displays the bolometric luminosity which is
clearly too flat after maximum \citep{clv00}. 
We suspect that both the peak is a
little too subluminous and that if the postmaximum fits better
reproduced the observed colors, we would have better agreement with
observed bolometric light curves after maximum. This shows that while
the density structure and composition of the outer layers of W7 do a
good job of reproducing the observed spectra, the structure of the
inner layers is not quite correct.

\section{Analysis}
\subsection{Identification of the 10500~\AA\ Feature\label{ir105}} 

\begin{figure}
\begin{center}
\psfig{file=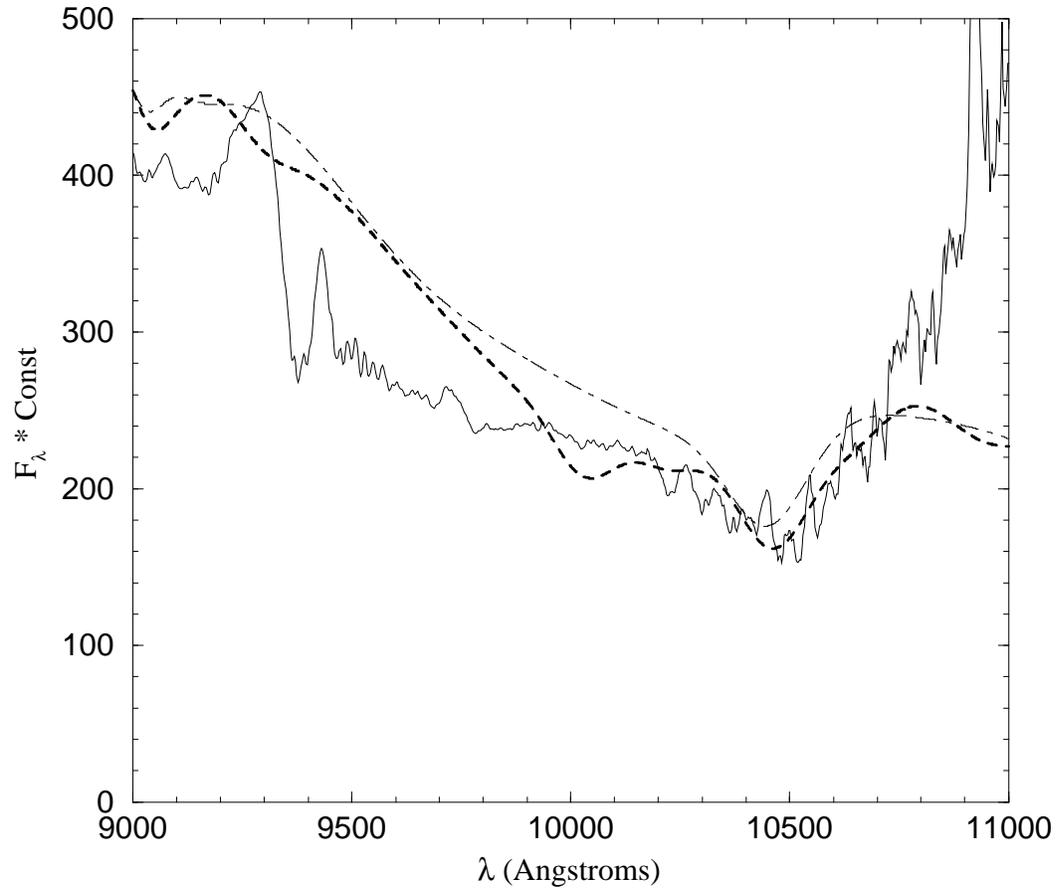,width=14cm}
\caption{Near-$IR$ spectrum on 20~March 1994 
\protect\citep[thin solid line][]{patat94D96}, best fit  
synthetic spectrum (thick dashed line), and best fit model using only
\ion{Mg}{2}\ line opacity. We believe that this firmly establishes the
identity of this feature.\label{fig:day19ir}}
\end{center}
\end{figure}

The identification of the absorption feature at 10500~\AA\ in the
spectra of \sneia\ has been the subject of debate. The primary
candidates are: \hei~$\lambda$10830, \ion{O}{1}~$\lambda$11287,
\mgii~$\lambda$10926, and \feii~$\lambda$10926. All of these, except
\hei, are found in W7.  We have treated all of these species in
NLTE. \citet{meikle94d91t} and \citet{mazz94d} find both \mgii\ and
\hei\ as equally probable.  \citet{hatano94D99} found the feature best
fit with \mgii.  \citet{whhs98}\ and \citet{mhwIR00}\ have used
delayed detonation models of C+O white dwarfs and found that the
\mgii\ provided a good fit and could be used as a diagnostic of the
outer edge of the burned material in \snia. We have also used a model,
W7, that does not contain helium and found that the feature fit
well. We have plotted the near-$IR$ spectrum of \snnfd\ on 20~March with
our best fit model and a diagnostic spectrum in
Figure~\ref{fig:day19ir}. The diagnostic spectrum is generated from
the converged atmosphere model and calculated with only \mgii\ line
opacity and no other line opacity. We can see that the 10500~\AA\
feature in all three spectra are of the same strength and at the same
wavelength.  This feature, which does not shift in wavelength in the
observations at earlier epochs \citep{meikle94d91t}, is clearly formed
by \mgii\ in our synthetic spectra of W7. This also shows that the
abundance of magnesium and the velocity range in W7 corresponds well
to that in \snnfd.

\subsection{Distance to NGC~4526 and the Luminosity of \snnfd}

\begin{figure}
\begin{center}
\psfig{file=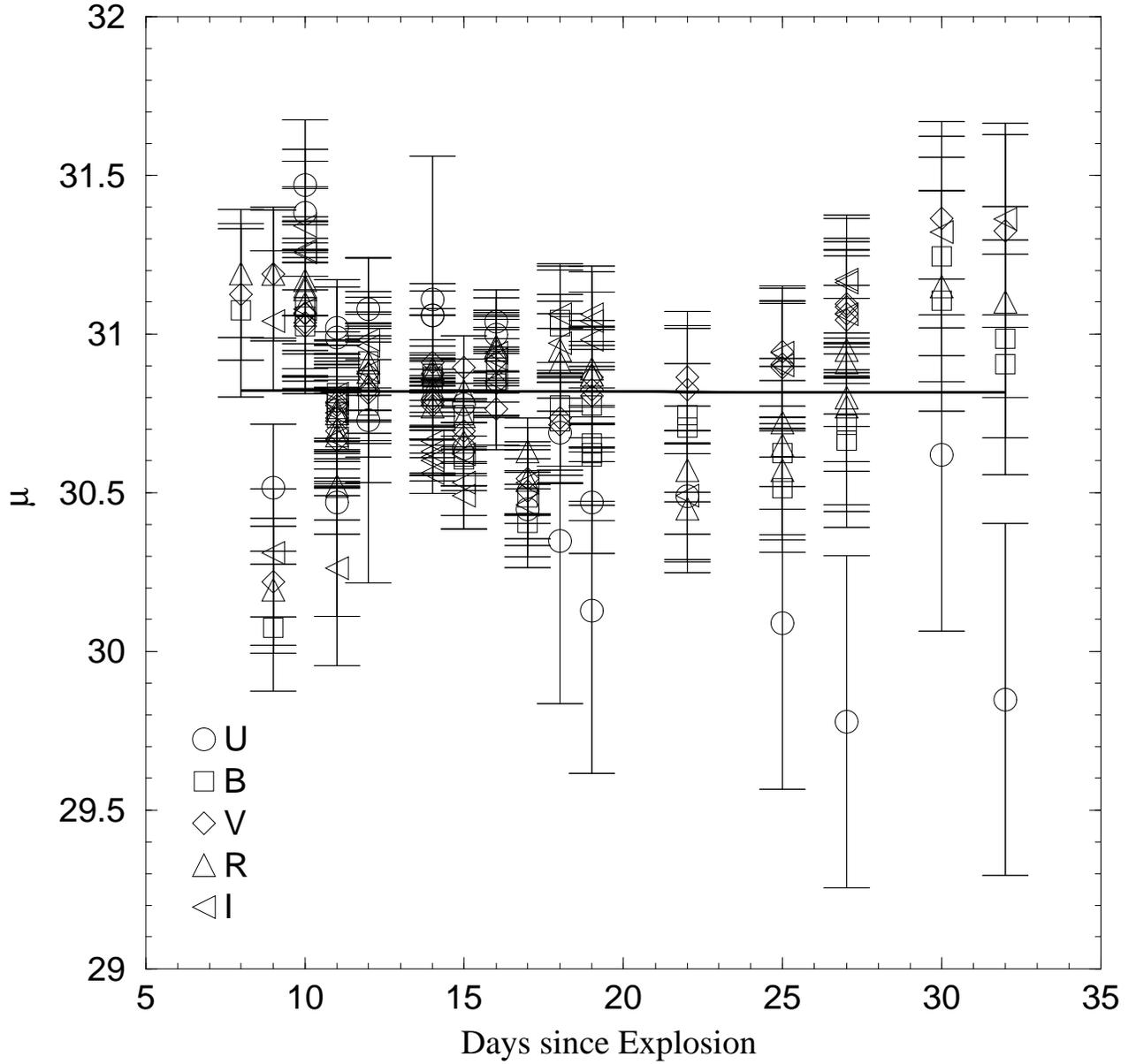,width=17cm,angle=-90}
\caption{Distance moduli calculated from available observed photometry and 
associated photometry from synthetic spectra. The error bars combine the
observers quoted errors and our estimate of the errors in our 
calculated luminosities. The solid line is a fit to the data which yields
$\mu = 30.8 \pm 0.3$, consistent within errors with the results of
\protect\citet{DR94D99} and \citet{hofsn94d}.  \label{fig:mu}}
\end{center}
\end{figure}

We have applied our synthetic photometry of \snnfd\ and the many
available observed photometric data to derive the distance to NGC~4526
using the Spectral-fitting Expanding Atmospheres Method (SEAM).
SEAM determines the distance to a supernova by fitting the spectra of a
supernova and deriving a distance modulus, $\mu$, from the synthetic
photometry \citep{b93j1,b93j2,b93j3,b93j4,b94i1,bsn99em00}.  We have
used all $U$, $B$, $V$, $R$, $I$ photometry from
\citet{richsn94d}, \citet{patat94D96}, \citet{meikle94d91t}, and
\citet{tsv94d94i} that correspond to 
the dates of the spectra we have fit in \S~\ref{spec}. We plot the
individual distance moduli obtained from each observation paired with
the appropriate synthetic spectrum in Figure~\ref{fig:mu}. The error
bars include the stated observational error and our estimate of the
fitting error of the model luminosity. For the $U$-band data of
\citet{richsn94d} we have used an error of 0.5 magnitudes to reflect
the systematic error suspected by the observers. We have computed an
error-weighted mean of the distance modulus and have found the value,
$\mu=30.8 \pm 0.3$, where 0.3 is the 1-$\sigma$ error.  
The
individual band averages range from $\mu = 30.71$ for the $I$-band to
$\mu = 30.94$ for the $U$-band.  The horizontal line in
Figure~\ref{fig:mu} is a weighted least-squares fit to the distance
modulus as a function of epoch. The fit varies by $\sim 0.01$
magnitudes over the range of epochs considered (the slope of the
fitted line is 0 to 2 parts in $10^{4}$). The luminosity of the
model atmosphere depends on the radius which is a function of
time. Small changes in the timing of the explosion will affect the
derived distance.  The spectra change slowly enough that a fit can be
nearly replicated with a model expanded to an earlier (later) epoch
with a lower (higher) luminosity. Each day error in the risetime
generates an error of $\sim 0.15$ magnitudes. There are also errors
that may arise from deficiencies in the chosen model and the other
inputs to the synthetic spectra. We derive the distance modulus to be $\mu
= 30.8 \pm 0.10$~(internal)~$\pm 0.20$~(timing). This result is 
consistent within errors with that of \citet{DR94D99} who find $\mu =
30.4 \pm 0.3$ 
using the globular cluster luminosity function, and with that of
\citet{hofsn94d} who used light curve calculations to obtain $\mu =
31.05 \pm 0.05$.

\subsection{Weak \ion{Ca}{2}~H+K and \ion{Na}{1}~D Features\label{cana}}

Two concerns about the synthetic spectra are the general weakness of the
\cahk\ feature in the models for the first 12 days after the
explosion, and the complete absence of the \naid\ feature, which
begins to appear about one week after maximum light. To overcome these
deficiencies we have attempted many modifications to the models. For the
early \cahk\ problem we have tried extending the models to higher
velocities. This does increase the strength of the \cair\ and \cahk\
features for the very early exponential models in
Figure~\ref{fig:early}, but does not affect the models which
correspond to the observed spectra, because at those later epochs the
line forming region has moved into the velocities of our normal W7
model.  Increasing the density of the C+O region by decreasing the
slope of the density decline did not have a significant effect. We also
tried increasing the deposition of \gamray\ energy, but as the problem
is over-ionization rather than under ionization, this did not improve
the \cahk\ feature shape. We have also tried extreme abundance
enrichment in calcium and sodium to make the corresponding features
stronger with inadequate results.  Thus, overall it will require
modifications to W7 to improve the lineshapes of the \cahk\
and \naid\ features. Since the problem is that e.g., Ca exists
primarily as \ion{Ca}{3} and we need to increase the \ion{Ca}{2}
population, this would require higher densities, since lower
temperatures would no longer fit the overall shape of the observed
spectrum. This could be an indication that non-spherical
effects, such as composition asymmetries and clumping are playing a role.

\subsection{How Good is W7?}

The synthetic spectra made with W7 fit the observed spectra of \snnfd\ 
quite well in the pre-maximum epoch, except for the above discussed
problem with \cahk. After the continuum `photosphere', $\tstd \sim 1$,
recedes into the Fe-peak element rich core the quality of the spectral
fits begins to break down. 
The other significant possibility is
that the particular density and abundance structure of W7 in the Fe-rich
core is not fully reflective of \sneia\ and \snnfd. Some neutron rich
isotopes have been shown to be overproduced in W7
\citep{iwamoto99,brach00}. 

The other deficiency is in the later pre-maximum spectra
where the 4000--6000~\AA\ flux is slightly low and certain temperature
sensitive features, such as \siii, indicate that the line forming
region should be hotter.  We found that these problems could be at
least partially solved by increasing the
luminosity, which seems to be required by the synthetic
photometry. However, these higher luminosity models fit the blue flux 
poorly. A model with more \nifs\ mixed into the partially
burned silicon and sulfur rich zones would raise the effective
temperature of the region that is above $\tstd = 1$, which could help
both the feature shape and the overall agreement with the inferred
maximum brightness of SNe~Ia.
This type of mixing would be expected in the turbulent
3-D models \citep{khokh01,LHWNK00,RHN99}.

\section{Conclusions}

The \snia\ C+O white dwarf deflagration model works reasonably
well in fitting the normal \sneia\ \snnfd. The fits are particularly
good for the week before maximum light. There are some modest problems
with the post-maximum spectra that are either due to differences
between the W7 Fe-rich core composition and structure and that of
\sneia, specifically \snnfd, or due to significant departures from LTE
in the temperature, ionization, and level populations of the Fe-peak
species in the core which requires a more detailed NLTE
treatment. From these results it seems reasonable to infer that the
outer parts of W7 reasonably well resemble the outer parts of SNe~Ia,
whereas the inner parts differ in density, composition, or nickel
mixing/clumping.  From the synthetic spectra we have computed synthetic
photometry. This synthetic photometry combined with the synthetic
specrta allow us to use the SEAM method to calculate the distance to
NGC~4526, which we find to be $\mu = 30.8 \pm 0.3$.
This is a very encouraging
demonstration of the 
potential of the SEAM method 
applied to \sneia.

Observationally, the fitting described in this paper
requires spectra that show a rise from the blue, through the peak of
the distribution, to the red tail.  Practically, this means spectra
covering the range 3100--8000~\AA, preferably extending to 10000~\AA\
or beyond. Near-$UV$ spectra and photometry are also needed for at least
some objects, so that we might understand what variability exists in
the near-$UV$. We hope that to have demonstrated that this data is accessible
with the current generation of instruments for \sneia\ out to $\mu
\sim 32$. Early discovery and monitoring are also important since we
can probe the outermost layers of \sneia\ only at these early times
and hopefully better constrain the risetimes of \sneia\ as a group and
individually.

We plan to consider additional explosion models, peculiar \sneia,
and infra-red spectra in more detail. Part of our efforts will
be to test the effects of more complete NLTE treatment of all
relevant ionic species in the opacity. We are also working to
include non-equilibrium temperature evolution effects through
the calculation of light curves.

\acknowledgements 
We thank Mario Hamuy for providing us with the
infrared filter and atmospheric transmission functions and for helpful
discussions on their proper interpretation and use. We also thank Alex
Filippenko, Ferdinando Patat, Nic Walton, Jim Lewis, and Peter 
Meikle for providing us with their observed spectra and photometry.  
This work was
supported in part by NSF grant AST-9731450, NASA grant NAG5-3505, and
an IBM SUR grant to the University of Oklahoma; and by NSF grant
AST-9720704, NASA ATP grant NAG 5-8425, and LTSA grant NAG 5-3619 to
the University of Georgia.  PHH was supported in part by the P\^ole
Scientifique de Mod\'elisation Num\'erique at ENS-Lyon. Some of the
calculations presented in this paper were performed at the San Diego
Supercomputer Center (SDSC), supported by the NSF, and at the National
Energy Research Supercomputer Center (NERSC), supported by the
U.S. DOE.  We thank both these institutions for a generous allocation
of computer time.


\begin{deluxetable}{crrrrrrrr}
\tablecolumns{9}
\tablewidth{0pc}
\tablecaption{Synthetic Photometry of Best-Fit Models\label{tab:phot}}
\tablehead{
\colhead{Epoch\tablenotemark{a}} & \colhead{U}   & \colhead{B}    & \colhead{V} &
\colhead{R}    & \colhead{I}   & \colhead{J}    & \colhead{H} & \colhead{K}}
\startdata
1\tablenotemark{b} & -11.57 & -12.13 & -13.03 & -13.30 & -13.19 &  -13.76 &  -13.80 &  -13.80 \\
2\tablenotemark{b} & -13.42 &  -13.82 &  -14.55 &  -14.66 &  -14.60 &  -15.08 &  -15.07 &  -14.94 \\
3\tablenotemark{b} &  -14.69 &  -14.97 &  -15.45 &  -15.46 &  -15.45 &  -15.77 &  -15.73 &  -15.51 \\
3\tablenotemark{c} &  -14.82 &  -14.90 &  -15.45 &  -15.45 &  -15.43 &  -15.74 &  -15.75 &  -15.50 \\
3 &  -15.01 &  -14.95 &  -15.41 &  -15.44 &  -15.52 &  -15.73 &  -15.73 &  -15.59 \\
4 & -15.89 &  -15.63 &  -16.02 &  -16.02 &  -16.11 &  -16.24 &  -16.23 &  -16.06 \\
5 & -16.48 &  -16.16 &  -16.45 &  -16.47 &  -16.56 &  -16.63 &  -16.60 &  -16.41 \\
6 &  -16.97 &  -16.55 &  -16.79 &  -16.83 &  -16.90 &  -16.95 &  -16.90 &  -16.67 \\
7 & -17.38 &  -16.88 &  -17.09 &  -17.15 &  -17.19 &  -17.21 &  -17.13 &  -16.86 \\
8 &  -17.58 &  -17.09 &  -17.29 &  -17.38 &  -17.42 &  -17.41 &  -17.31 &  -16.96 \\
9 &  -18.39 &  -17.68 &  -17.69 &  -17.79 &  -17.78 &  -17.67 &  -17.45 &  -16.90 \\
10 &  -18.87 &  -18.04 &  -17.98 &  -18.17 &  -18.19 &  -18.14 &  -18.03 &  -17.82 \\
11 &  -18.84 &  -18.12 &  -18.03 &  -18.14 &  -18.06 &  -17.85 &  -17.63 &  -16.98 \\
12 &  -19.26 &  -18.49 &  -18.39 &  -18.55 &  -18.47 &  -18.25 &  -18.10 &  -17.79 \\
14 &  -19.71 &  -18.84 &  -18.69 &  -18.76 &  -18.43 &  -18.15 &  -17.93 &  -17.54 \\
15 & -19.43 &  -18.76 &  -18.71 &  -18.76 &  -18.44 &  -18.02 &  -17.78 &  -17.30 \\
16 &  -19.88 &  -19.10 &  -18.95 &  -19.12 &  -18.88 &  -18.42 &  -18.23 &  -17.94 \\
17 &  -19.43 &  -18.76 &  -18.71 &  -18.76 &  -18.44 &  -18.02 &  -17.78 &  -17.30 \\
18 & -19.68 &  -19.11 &  -18.99 &  -19.15 &  -18.94 &  -18.55 &  -18.41 &  -18.12 \\
19 & -19.43 &  -19.02 &  -19.09 &  -19.10 &  -18.93 &  -18.69 &  -18.64 &  -18.44 \\
22 &  -19.36 &  -19.06 &  -19.11 &  -18.73 &  -18.33 &  -18.19 &  -18.11 &  -17.39 \\
25 &  -18.55 &  -18.74 &  -19.11 &  -18.75 &  -18.57 &  -18.42 &  -18.60 &  -17.67 \\
27 & -18.19 &  -18.66 &  -19.13 &  -18.76 &  -18.62 &  -18.44 &  -18.78 &  -17.72 \\
30 &  -18.57 &  -18.75 &  -19.18 &  -18.74 &  -18.64 &  -18.23 &  -18.82 &  -17.15 \\
32 &  -17.57 &  -18.27 &  -19.02 &  -18.59 &  -18.68 &  -18.36 &  -19.16 &  -17.24 \\
\enddata
\tablenotetext{a}{Days after explosion.}
\tablenotetext{b}{Exponential density profile of C+O composition}
\tablenotetext{c}{Exponential density profile with C+O composition.\\
 Maximum velocity, $\sim 30000$~km~s$^{-1}$.}
\tablenotetext{d}{As can be seen the value of $M_{B_{\rm max}}$, the peak
luminosity is a bit low ($\sim 0.3$~mag), which reflects some
shortcomings of the intermediate velocity parts of the W7 model.}
\end{deluxetable}

\clearpage
\begin{deluxetable}{rrrrrrrrrrrrr}  
\tablecolumns{13}  
\tablewidth{0pc}  
\tablecaption{Color Comparison\label{tab:color}}
\tablehead{  
\colhead{}  & \colhead{}  &  \multicolumn{2}{c}{U-B} &   \colhead{}   &  
\multicolumn{2}{c}{B-V} &   \colhead{}   &  
\multicolumn{2}{c}{V-R} &   \colhead{}   &  \multicolumn{2}{c}{R-I} \\
\cline{3-4} \cline{6-7} \cline{9-10} \cline{12-13} \\  
\colhead{Epoch\tablenotemark{a}} & \colhead{} &
\colhead{syn}   & \colhead{obs}    & \colhead{} &  
\colhead{syn}   & \colhead{obs}    & \colhead{} &  
\colhead{syn}   & \colhead{obs}    & \colhead{} &  
\colhead{syn}   & \colhead{obs}   }
\startdata  
8\richm & & -.45 & \nodata & & .27 & .21 & & .13 & .07 & & .09 & \nodata \\
9\jktm & &  -.67 & -.22 & &    .07 & -.07 & &    .15 & .17 & &    .04 & -.08 \\
9\intm & & & \nodata & & & \nodata & & & .15 & & & .12 \\
10\richm & &  -.78 & \nodata& &   .01 &-.05 & &    .23 & .12 & &    .07 &-.04 \\
10\patm & & & -.37 & & & -.01 & & &.16 & & &-.13 \\
10\jktm & & & -.43 & & & -.04 & & &.13 & & &-.02 \\
10\intm & & & \nodata & & & \nodata & & & .16 & & & .09 \\
11\richm & & -.67 & -.97 & &  -.04 & .00 & &   .16 & .14 & &  -.03 & -.06 \\
11\patm & & & -.47 & & & -.02 & & &.23 & & &-.14 \\
11\jktm & & & -.47 & & & .00 & & & .18 & & & .06 \\
11\intm & & & \nodata & & & \nodata & & & .30 & & & .23 \\
12\richm & & -.73 & -.85 & &   -.04 & -.01 & &    .21 & .12 & &   -.04 & -.08\\ 
12\patm & & & -.56 & & & .00 & & &.24 & & &-.17 \\
14\richm & &  -.83 & -.62 & &   -.09 & -.01 & &    .11 & .08 & &   -.27 & -.07 \\ 
14\patm & & & -.62 & & & -.03 & & & .14 & & &-.12 \\
14\jktm & & & -.58 & & & -.10 & & & .10 & & & -.05 \\
14\intm & & & \nodata & & & \nodata & & & .15 & & & -.03 \\
15\richm & &  -.63 & \nodata & &  .02 & -.06 & &    .09 & .03 & &   -.27 & -.02 \\
15\patm & & & -.60 & & & -.07 & & & .13 & & &-.14 \\
15\jktm & & & -.59 & & & -.14 & & & .17 & & & -.07 \\
16\richm & &  -.74 & \nodata & &   -.08 & .00 & &    .21 & .03 & &   -.18 & -.09 \\
16\patm & & & -.58 & & & -.09 & & & .10 & & &-.17 \\
16\jktm & & & -.58 & & & -.12 & & & .17 & & & -.15 \\
17\richm & &  -.76 & -.95 & &   -.09 & .00 & &    .19 & -.03 & &   -.21 & -.16 \\ 
17\patm & & & -.58 & & & -.09 & & & .06 & & &-.26 \\
18\richm & &  -.53 & -.95 & &   -.06 & .00 & &    .21 & -.03 & &   -.16 & -.18 \\ 
18\patm & & & -.56 & & & -.07 & & & .03 & & &-.31 \\
19\richm & & -.36 & -.89 & &    .12 & -.03 & &    .06 & -.03 & &   -.12 & -.27\\
19\patm & & & -.51 & & & -.08 & & & .01 & & &-.31 \\
19\intm & & & \nodata & & & -.11 & & & .05 & & & -.23 \\
22 & &   -.25 & & &    .11 & & &   -.34 & & &   -.34 & \\
25\richm & &  .24 & -.30 & &    .42 & .11 & &   -.31 & -.09 & &   -.13 & -.35\\
25\patm & & & \nodata & & & .05 & & & .01 & & & \nodata \\
 & & & \nodata & & & \nodata & & & -.06 & & & -.38 \\
27\richm & &  .51 & -.42 & &    .53 & .16 & &   -.32 & -.15 & &   -.09 & -.35\\
27\patm & & & \nodata & & & .15 & & & -.05 & & & \nodata \\
 & & & \nodata & & & \nodata & & & -.06 & & & -.35 \\
30\richm & &  .22 & -.26 & &    .49 & .23 & &   -.39 & -.18 & &   -.05 & -.22\\
30\patm & & & \nodata & & & \nodata & & & \nodata & & & -.37 \\
32\richm & &  .74 & -.31 & &    .81 & .39 & &   -.38 & -.16 & &    .14 & -.12\\
32\patm & & & -.05 & & & .27 & & & \nodata & & & \nodata \\
\enddata
\tablenotetext{a}{Days after explosion.}
\tablenotetext{b}{Data from \citet{richsn94d}.}
\tablenotetext{c}{Data from \citet{patat94D96}.}
\tablenotetext{d}{Data from \citet{meikle94d91t}\ using Jacobus Kapteyn Telescope.}
\tablenotetext{e}{Data from \citet{meikle94d91t}\ using Issac Newton Telescope.}
\end{deluxetable}
\clearpage

\begin{deluxetable}{rrrrrrrr}  
\tablecolumns{8}  
\tablewidth{0pc}  
\tablecaption{HST ACS Synthetic Photometry\label{tab:uv}}
\tablehead{  
\colhead{}    &  \multicolumn{3}{c}{Magnitude} &   \colhead{}   &  
\multicolumn{3}{c}{Count Rate} \\  
\cline{2-4} \cline{6-8} \\  
\colhead{Epoch\tablenotemark{a}} & \colhead{F330W}   & \colhead{F250W}    & \colhead{F220W} &  
\colhead{} & \colhead{F330W}   & \colhead{F250W}    & \colhead{F220W} }  
\startdata 
   1\tablenotemark{b} &  -10.47 &   -8.86 &   -6.73 & &      5.73 &       .87 &       .07 \\
   2\tablenotemark{b} &  -12.46 &  -11.18 &   -8.96 & &     35.56 &      7.35 &       .58 \\
   3\tablenotemark{b} &  -14.24 &  -12.89 &  -11.16 & &    183.97 &     35.47 &      4.42 \\
   3\tablenotemark{c} &  -14.63 &  -13.28 &  -11.50 & &    262.76 &     50.72 &      6.02 \\
   3 &  -13.61 &  -12.26 &  -10.02 & &    102.93 &     19.81 &      1.55 \\
   4 &  -15.75 &  -14.30 &  -12.47 & &    740.02 &    129.45 &     14.77 \\
   5 &  -16.36 &  -14.92 &  -12.81 & &   1298.30 &    230.24 &     20.20 \\
   6 &  -16.93 &  -15.54 &  -13.37 & &   2195.12 &    407.04 &     33.79 \\
   7 &  -17.41 &  -16.16 &  -14.07 & &   3422.05 &    722.22 &     64.44 \\
   8 &  -17.54 &  -16.03 &  -14.01 & &   3839.12 &    638.16 &     60.60 \\
   9 &  -18.52 &  -17.11 &  -14.83 & &   9466.57 &   1724.80 &    128.85 \\
  10 &  -19.22 &  -18.35 &  -16.65 & &  18108.65 &   5403.09 &    690.96 \\
  11 &  -19.07 &  -18.00 &  -16.12 & &  15671.64 &   3900.93 &    423.30 \\
  12 &  -19.57 &  -18.58 &  -16.75 & &  25029.80 &   6712.84 &    757.36 \\
  14 &  -20.11 &  -19.14 &  -17.31 & &  40829.55 &  11160.32 &   1271.70 \\
  15 &  -19.57 &  -18.26 &  -16.54 & &  24985.13 &   4960.63 &    625.50 \\
  16 &  -20.22 &  -19.17 &  -17.30 & &  45226.13 &  11464.97 &   1257.49 \\
  17 &  -20.34 &  -19.22 &  -17.18 & &  50561.80 &  12045.89 &   1129.03 \\
  18 &  -19.86 &  -18.80 &  -16.93 & &  32516.13 &   8165.91 &    897.51 \\
  19 &  -19.38 &  -17.65 &  -15.48 & &  20898.99 &   2826.09 &    236.23 \\
  22 &  -19.00 &  -17.05 &  -14.85 & &  14799.15 &   1634.24 &    131.61 \\
  25 &  -17.68 &  -16.03 &  -13.76 & &   4364.39 &    635.58 &     48.50 \\
  27 &  -17.00 &  -15.65 &  -13.42 & &   2328.59 &    450.16 &     35.16 \\
  30 &  -17.92 &  -16.25 &  -13.96 & &   5445.12 &    783.58 &     58.06 \\
  32 &  -16.37 &  -15.22 &  -12.94 & &   1310.45 &    302.33 &     22.61 \\
\enddata
\tablenotetext{a}{Days after explosion.}
\tablenotetext{b}{Exponential density profile of C+O composition}
\tablenotetext{c}{Exponential density profile with C+O composition.\\
 Maximum velocity, $\sim 30000$~km~s$^{-1}$.}
\end{deluxetable}
\clearpage


\clearpage

\clearpage

\clearpage

\clearpage

\clearpage

\clearpage

\clearpage

\clearpage

\clearpage

\clearpage

\end{document}